\documentclass[10pt,a4paper]{article}
\usepackage{jcappub}

\allowdisplaybreaks

\usepackage{ifthen}
\usepackage{graphicx}
\usepackage{dcolumn}
\usepackage{bm}
\usepackage{color}
\usepackage{amsmath}
\usepackage{amssymb}
\usepackage{subfigure}

\newcommand{\vp}{\mathbf{p}}
\newcommand{\vq}{\mathbf{q}}
\newcommand{\vs}{\mathbf{s}}
\newcommand{\vx}{\mathbf{x}}

\newcommand{\vk}{\mathbf{k}}

\newcommand{\himpc}{{\hbox {$~h^{-1}$}{\rm ~Mpc}}}
\newcommand{\hmpci}{{\hbox {$~h{\rm ~Mpc}^{-1}$}}}

\newcommand{\be}{\begin{equation}}
\newcommand{\ee}{\end{equation}}

\begin{document}

\title{Distribution function approach to redshift space distortions. Part III: halos and galaxies}

\author[a]{Teppei Okumura,} \emailAdd{teppei@ewha.ac.kr}
\author[a,b,c]{Uro{\v s} Seljak,} \emailAdd{useljak@berkeley.edu}
\author[d]{and Vincent Desjacques} \emailAdd{dvince@physik.uzh.ch}

\affiliation[a]{Institute for the Early Universe, Ewha Womans
  University, Seoul 120-750, S. Korea}

\affiliation[b]{Department of Physics, Department of Astronomy, and Lawrence Berkeley National
  Laboratory, University of California, Berkeley, California 94720,
  USA} 

\affiliation[c]{Institute of Theoretical Physics, University of Zurich, 8057 Zurich, Switzerland}

\affiliation[d]{D\'epartement de Physique Th\'eorique and Center for Astroparticle Physics (CAP), 
Universit\'e de Gen\`eve, 1211 Gen\`eve, Switzerland}

\abstract{It was recently shown that the power spectrum in redshift
  space can be written as a sum of cross-power spectra between number
  weighted velocity moments, of which the lowest are density and
  momentum density. We investigate numerically the properties of these
  power spectra for simulated galaxies and dark matter halos and
  compare them to the dark matter power spectra, generalizing the
  concept of the bias in density-density power spectra.  Because all
  of the quantities are number weighted this approach is well defined
  even for sparse systems such as massive halos. This contrasts to the
  previous approaches to RSD where velocity correlations have been
  explored, but velocity field is a poorly defined concept for sparse
  systems.  We find that the number density weighting leads to a
  strong scale dependence of the bias terms for momentum density
  auto-correlation and cross-correlation with density.  This trend
  becomes more significant for the more biased halos and leads to an
  enhancement of RSD power relative to the linear
  theory. Fingers-of-god effects, which in this formalism come from
  the correlations of the higher order moments beyond the momentum
  density, lead to smoothing of the power spectrum and can reduce this
  enhancement of power from the scale dependent bias, but are
  relatively small for halos with no small scale velocity dispersion.
  In comparison, for a more realistic galaxy sample with satellites
  the small scale velocity dispersion generated by satellite motions
  inside the halos leads to a larger power suppression on small
  scales, but this depends on the satellite fraction and on the
  details of how the satellites are distributed inside the halo.  We
  investigate several statistics such as the two-dimensional power
  spectrum $P(k,\mu)$, where $\mu$ is the angle between the Fourier
  mode and line of sight, its multipole moments, its powers of
  $\mu^2$, and configuration space statistics.  Overall we find that
  the nonlinear effects in realistic galaxy samples such as luminous
  red galaxies affect the redshift space clustering on very large
  scales: for example, the quadrupole moment is affected by 10\% for
  $k<0.1\hmpci$, which means that these effects need to be understood
  if we want to extract cosmological information from the redshift
  space distortions.  } 
%\pacs{98.80, 98.65, 98.62}
\arxivnumber{1206.XXXX}

\maketitle

%%%%%%%%%%%%%%%%%%%%%%%%%%%%%%%%%%%%%%%%%%%%%%
% Section 1 Introduction
%%%%%%%%%%%%%%%%%%%%%%%%%%%%%%%%%%%%%%%%%%%%%%
\section{Introduction}\label{sec:intro}
Galaxy redshift surveys are one of the most powerful tools to probe
cosmological models \cite{Peebles:1980}.  Galaxy distribution in
redshift surveys is distorted through the Doppler shift by peculiar
velocities of galaxies along the line of sight.  Thus the measured
redshift of the galaxy provide not only the information of the radial
distance but also that of the radial velocity.  This effect, so called
redshift-space distortions (RSD), induces anisotropies in the galaxy
clustering and allows one to measure the amplitude of density
fluctuations times the rate of growth of structure on large scales,
$f=d\ln D/d\ln a$, with $D$ the linear growth factor
\cite{Kaiser:1987, Hamilton:1998}.  Cosmological models in different
gravity theories can have a different value of $f$, thus RSD are a
promising tool to investigate gravity theories
\cite[e.g.,][]{Linder:2005, Jain:2008, Guzzo:2008, Song:2009}.  So far
RSD have been analyzed in many galaxy surveys to determine the
cosmological models \citep[e.g.,][]{Peacock:2001, Zehavi:2002,
  Hawkins:2003, Tegmark:2004, Tegmark:2006, Ross:2007, Guzzo:2008,
  Okumura:2008, Cabre:2009, Blake:2011}.  However, it was shown by
\cite{Tinker:2006, Desjacques:2010a, Okumura:2011, Jennings:2011,
  Kwan:2011} that the parameter reconstructed from the redshift-space
distortions can have scale dependent bias, which indicates a breakdown
of the linear theory predictions. These effects show up on relatively
large scales, suggesting one must go beyond the linear theory in the
analysis of RSD.

Given the high precision of the future surveys, correspondingly more
accurate theoretical predictions become essential for their
interpretation.  Recently there have been many studies to predict the
power spectrum in nonlinear regime beyond the framework of the
standard perturbation theory (SPT) \cite{Bernardeau:2002, Crocce:2006,
  Crocce:2006b, Matarrese:2007, McDonald:2007, Valageas:2007,
  Taruya:2008}.  Similarly, initial RSD work was based on the lowest
order SPT \cite{Heavens:1998, Scoccimarro:1999, Bharadwaj:2001,
  Pandey:2005}.  However, as pointed out by \cite{Scoccimarro:1999,
  Scoccimarro:2004}, SPT in redshift space breaks down at larger
scales than in real space because of nonlinear redshift distortion
effects. Sometimes this is attributed to the so-called Fingers-of-God
(FoG) effect \cite{Jackson:1972}. However, we will argue that a more
important effect is the scale dependent biasing.  Recent development
using more sophisticated perturbation methods applicable to the
redshift-space power spectrum includes \cite{Matsubara:2008,
  Taruya:2010, Valageas:2011, Seljak:2011}.

There is another issue that needs to be taken into account to achieve
accurate theoretical predictions of RSDs.  Galaxies, or dark matter
halos within which all galaxies are expected to form, are a biased
tracer of dark matter, as their clustering strength is typically
enhanced relative to the dark matter, which is known as biasing
\cite{Kaiser:1984, Bardeen:1986}.  Because of the existence of the
bias, RSD on linear scales in galaxy surveys allow us to measure the
linear RSD parameter $\beta=f/b$, where $b$ is the bias parameter.
This can be combined with the auto-correlation of galaxies to
eliminate the bias and measure $fA$, where $A$ is the amplitude of
density fluctuations (often parameterized with $\sigma_8$).  Recently
it was shown by \cite{Okumura:2011} using linear theory that the RSD
parameters reconstructed from the clustering of halos have strong
halo-mass and scale dependence even on large scales.  It is, however,
not trivial to incorporate the bias into an analytical framework of
nonlinear perturbation theory, since the galaxy formation is a highly
nonlinear process.  One attempt was presented by
\cite{Matsubara:2008a, Sato:2011} in redshift space using the
Lagrangian perturbation theory. However, \cite{Reid:2011a} show that
the formula of \cite{Matsubara:2008a} seriously fails to predict the
quadrupole and hexadecapole moments of the redshift-space correlation
function even on very large scales.  \cite{Nishimichi:2011} extended
the formalism of \cite{Taruya:2010} by combining with a simple halo
bias scheme and tested it using dark matter halo catalogs from
$N$-body simulations (see also \cite{de-la-Torre:2012}).  There are
also several studies to attempt to eliminate such nuisance effects by
combining RSD to the other measurement.  \cite{Zhang:2007} proposed a
method to eliminate the uncertainty of the galaxy biasing by combining
weak gravitational lensing, galaxy clustering and RSD (see
\cite{Reyes:2010} for the observational result).  Similarly,
\cite{Hikage:2012} developed an approach for using galaxy-galaxy weak
lensing to model the FoG effect in RSD measurements.

A recent paper \cite{Seljak:2011} (Part I in the series of papers
studying RSD) has developed a phase space distribution function
approach to RSD where the redshift-space density can be written as a
sum over mass or number weighted moments of radial velocity, which are
integrals of powers of velocity over the momentum part of the phase
space distribution function. The corresponding RSD power spectrum can
be written as a sum over auto and cross-correlators of these moments.
In \cite{Okumura:2012} (Part II) we analyzed a large set of $N$-body
simulations to test how accurately this formalism predicts the true
power spectrum of dark matter as a function of terms included.  The
expansion was compared to the Legendre moments, the monopole,
quadrupole and hexadecapole moments.  These comparisons revealed that
the expansion is accurate within a few percent up to $k\simeq
0.15(1+z)\hmpci$ if the corrections up to the 6th order are taken into
account.  We also presented a resummation of some of the terms into a
power suppression factors called the FoG kernel.  This FoG model has
validity comparable or better than the 6th order summations and
predicts the monopole power spectrum with a few percent accuracy up to
$k\simeq 0.4\hmpci$ at $z=0$ and for $k<1\hmpci$ at $z=0.5$ and 1.
\cite{Seljak:2011} has also shown that the moments can be decomposed
into helicity eigenstates, which are eigenmodes under rotation around
direction of $\vk$ vector. Only equal helicity eigenstates correlate,
leading to a specific angular structure of the correlators. This
analysis shows that if one expands the series into powers of
$\mu^{2j}$, a finite number of terms contribute at each (finite)
order.  This suggests that RSD can be better understood in terms of
this expansion rather than the Legendre moments usually used.  Using
the angular decomposition the individual terms for the coefficients of
$\mu^{2j}$ for the dark matter power spectrum were determined.
Detailed comparison of the numerical results to perturbation theory
predictions will be made in \cite{Vlah:2012} (Part IV).

This paper is Part III in this series.  Because RSD is described by
number-weighted velocity moment correlators, there is a particular
advantage when we analyze galaxies and dark matter halos, since for
sparse systems volume-weighted velocity moments cannot be easily
defined.  In this paper we test our formalism to describe the
redshift-space power spectrum of galaxies and halos in nonlinear
regime using a large set of cosmological $N$-body simulations, as well
as present the individual terms of expansion for comparison against
each other, as an extension of the analysis of the dark matter power
spectrum in Part II \cite{Okumura:2012}.  The structure of this paper
is as follows.  In section \ref{sec:theory} we briefly review the
distribution function approach to RSD and extend it for biased objects
such as halos.  Section \ref{sec:sim} describes the $N$-body
simulations and how to measure the two-point statistics used in this
paper.  Section \ref{sec:analysis} presents the power spectra of the
number-weighted velocity moments and their contributions to the full
2-d spectrum in redshift space. We also discuss in detail properties
of biasing of halos and galaxies using these power spectra.  In
section \ref{sec:mu} we apply our formalism to powers of $\mu$
expansion, showing individual contributions to $\mu^0$, $\mu^2$,
$\cdots$, $\mu^8$ terms in $P^{s}$.  In section \ref{sec:tpcf} we
present the redshift-space correlation functions of halos and galaxies
and compare to the power spectrum analysis.  Section
\ref{sec:conclusion} is devoted to conclusions of this paper.

%%%%%%%%%%%%%%%%%%%%%%%%%%%%%%%%%%%%%%%%%%%%%%
% Section 2 RSD
%%%%%%%%%%%%%%%%%%%%%%%%%%%%%%%%%%%%%%%%%%%%%%

\section{Redshift-space distortions from the distribution function}
\label{sec:theory}
Throughout this paper we adopt a phase-space distribution function
approach to describe redshift-space distortions, proposed by
\cite{McDonald:2011, Seljak:2011}.  This approach was tested to dark
matter simulations by \cite{Okumura:2012}, and a similar discussion is
applicable to dark matter halos and galaxies.  The exact evolution of
collisionless particles is described by the Vlasov equation
\cite{Peebles:1980}. We thus start from the distribution function of
particles $f(\vx,\vq,t)$ at phase-space position $(\vx,\vq)$ in order
to derive the perturbative redshift-space distortions.  Here $\vx$ is
the comoving position and $\vq=\vp/a$ is the comoving momentum ($\vp$
is the proper momentum).  The density field in redshift space is
related to moments of distribution function as
\begin{equation}
  \delta_s^m(\vk)=
  \sum_{L=0}\frac{1}{L!}
  \left(\frac{i k_\parallel}{\cal H}\right)^L T_\parallel^{L,m}(\vk) ~,
  \label{eq:deltak}
\end{equation}
where the superscript $m$ denotes quantities for dark matter and
${\cal H}=aH$ where $H$ is the Hubble parameter.
$T_\parallel^{L,m}(\vk)$ is the Fourier transform of
$T_\parallel^{L,m}(\vx)$, defined as
\be
  T_\parallel^{L,m}(\vx)={m_p \over \bar{\rho}}
  ~ \int d^3\vq~ f\left(\vx,\vq\right)
  u_\parallel^{L}= \left\langle \left(1+\delta^m(\vx)\right) u_\parallel^{L}(\vx) \right\rangle_{\vx}, \label{eq:q_def}
\ee
where $u_\parallel$ is the radial peculiar velocity,\footnote{Unlike
  the definition here, in our previous paper \cite{Okumura:2012} the
  velocity $u_\parallel$ and the velocity moments $T_{\parallel}^L$
  were defined in comoving coordinates, thus $H$ instead of ${\cal H}$
  was used in the formalism. These two expressions are essentially the
  same.  However, in \cite{Okumura:2012} there was an obvious typo in
  the definition of $u_\parallel$: it should have been
  $am_pu_\parallel=q_\parallel$.}
$m_pu_\parallel=q_\parallel = \vq\cdot \hat{r}$, $m_p$ is the particle
mass, $\hat{r}$ is the unit vector pointing along the observer's line
of sight and $\bar{\rho}$ is the mean mass density.  The power
spectrum in redshift space is then given by \cite{Seljak:2011,
  Okumura:2012},
\begin{equation}
  P^{s}_{mm}(\vk)=\sum_{L=0}^{\infty}\sum_{L'=0}^{\infty}\frac{\left(-1\right)^{L'}}{L!~L'!}
  \left(\frac{i k\mu}{\cal H}\right)^{L+L'} P^{mm}_{LL'}(\vk) ~,  \label{eq:p_ss_dm}
\end{equation}
where $k_{||}/k=\cos \theta=\mu$ and
$P^{mm}_{LL'}(\vk)\delta(\vk-\vk')=\langle T_{\parallel}^{L,m}(\vk)
(T_\parallel^{*L',m}(\vk')) \rangle$.

The model of the power spectrum for mass presented above can be
extended to biased objects such as dark matter halos and galaxies
without any assumption.  It is given by simply replacing the
superscript and subscript $m$ in equation (\ref{eq:p_ss_dm}) by $h$,
denoting quantities for halos, as
\begin{eqnarray}
  P^{s}_{hh}(\vk)
  &=&\sum_{L=0}^{\infty}\sum_{L'=0}^{\infty}\frac{\left(-1\right)^{L'}}{L!~L'!}
  \left(\frac{i k\mu}{\cal H}\right)^{L+L'} P^{hh}_{LL'}(\vk) \nonumber \\
 & =&\sum_{L=0}^{\infty}\frac{1}{L!^2}\left(\frac{ k\mu}{\cal H}\right)^{2L} P^{hh}_{LL}(\vk) +
  2\sum_{L=0}^{\infty}\sum_{L'>L}\frac{\left(-1\right)^{L'}}{L!~L'!}
  \left(\frac{i k\mu}{\cal H}\right)^{L+L'} P^{hh}_{LL'}(\vk) \label{eq:p_ss_halo} ~.
\end{eqnarray}
We will sometimes omit the super/subscript $m$ and $h$ in the
following when a given equation holds for both dark matter and halos.

It is useful to compare this to Kaiser's linear theory prediction
\cite{Kaiser:1987, Scoccimarro:2004}.  If we approximate the
expression with the lowest 3 terms $P_{00}$, $P_{01}$ and $P_{11}$ and
assume standard linear theory, we obtain the linear Kaiser formula, as
\be  P^{s}_{hh,{\rm Kaiser}}(\vk)=\left( b+f\mu^2 \right)^2 P_{00,{\rm lin}}^{mm}(k); \ \ {\rm linear}, \label{eq:kaiser}
\ee
where $b$ is the bias parameter (see section \ref{sec:bias} below) and
$f=d\ln{D}/d\ln{a}$ with $D$ the growth factor.  If the nonlinear
corrections for these terms are taken into account,
\begin{eqnarray}  
P^{s}_{hh,{\rm Kaiser}}(\vk) &=& P^{hh}_{00}(k) + 2f\mu^2\left(\frac{ik}{{\cal H}\mu f}\right)P^{hh}_{01}(k)+f^2\mu^4\left(\frac{k}{{\cal H}\mu f}\right)^2 P^{hh}_{11}(k); \ \   {\rm nonlinear}  \\
&=& b^2P^{mm}_{00}(k) + 2bf\mu^2\left(\frac{ik}{{\cal H}\mu f}\right)P^{mm}_{01}(k)+f^2\mu^4 \left(\frac{k}{{\cal H}\mu f}\right)^2 P^{mm}_{11}(k), \label{eq:nl_kaiser}  
\end{eqnarray}
i.e., it is given by the lowest 3 terms $P_{00}$, $P_{01}$ and
$P_{11}$.  Since these terms have nonlinear corrections, we call this
approximation the nonlinear Kaiser order approximation.  Replacing
these lowest 3 moments with the standard linear theory we obtain the
original linear Kaiser model of equation (\ref{eq:kaiser}), and
\be
P^{mm}_{00, {\rm lin}}(\vk) = \left( \frac{ik}{{\cal H}\mu f} \right)P^{mm}_{01, {\rm lin}}(\vk) = \left( \frac{ik}{{\cal H}\mu f} \right)^2 P^{mm}_{11, {\rm lin}}(\vk). \label{eq:pll_kaiser}
\ee
Here we want to view this series simply as a series in $k_\parallel$,
investigating the convergence as more terms are added.

Note that the calculations never require anything but simple power
spectra of number-weighted powers of velocity to be computed from the
simulations.  These number-weighted quantities are well-defined even
for sparse biased systems such as halos or galaxies.  The order of
$k_\parallel=k\mu$ needed for convergence to a given level of accuracy
will inevitably increase as one goes to increasingly small scales,
with the whole expansion eventually breaking down once $k\mu
\sigma/{\cal H}>1$, where $\sigma$ is a typical velocity of the system.

\subsection{Angular decomposition and relation to Legendre multipole moments}\label{sec:th_angular}

By performing helicity decomposition \cite{Seljak:2011} show that the
power spectrum can be written as
\be
P_{LL'}(\vk)=\sum_{(l=L,L-2,..)}\sum_{(l'=L',L'-2,..;\; l'\ge l)}\sum_{n=0}^{l}P^{L,L',n}_{l,l'}(k)P_l^n(\mu)P_{l'}^n(\mu) ,
\label{eq:pll}
\ee
where $P_l^n(\mu)$ are the associated Legendre polynomials, which
determine the angular dependence of the spherical harmonics. There are
5 numbers that describe these objects: $L$ and $L'$ describe the power
of two velocity moments we are correlating, $l$, $l'$ describe the
rank of the object, for example $l=1$ is rank-1, which is a 3-d
vector, $l=2$ is a 3-d tensor etc.  Finally, $n$ is the helicity
eigennumber,\footnote{$m$ is usually used in place of $n$, but here the
  superscript $m$ is used to describe quantities for dark matter.}
which ranges between 0 and $l$ ($l \le l'$). Only equal helicity
components of expansion have a non-vanishing correlator.  There is a
close relation between the order of the moments and their angular
dependence.  The lowest contribution in powers of $\mu$ to $P^{s}(k)$
is $\mu^{L+L'}$ if $L+L'$ is even or $\mu^{L+L'+1}$ if $L+L'$ is odd,
and the highest is $\mu^{2(L+L')}$. Thus for $P_{00}(\vk)$ the only
angular term is isotropic term ($\mu^0$), for $P_{01}(\vk)$ the only
angular term is $\mu^2$, $P_{11}(\vk)$ and $P_{02}(\vk)$ contain both
$\mu^2$ and $\mu^4$ etc.  Note that only even powers of $\mu$ enter in
the final expression, as required by the symmetry.  We can thus write
\begin{equation}
  P^{s}(\vk)=\sum_{L=0}^{\infty}\frac{1}{L!^2}\left(\frac{ k}{\cal H}\right)^{2L} \sum_{j=2L}^{4L}P^{(j)}_{LL}(\vk)\mu^{j} +
  2\sum_{L=0}^{\infty}\sum_{L'>L}\frac{\left(-1\right)^{L}}{L!~L'!}
  \left(\frac{i k}{\cal H}\right)^{L+L'} \sum_{j=(L+L') {\rm or} (L+L'+1)}^{2(L+L')}P^{(j)}_{LL'}(\vk)\mu^{j} \label{eq:p_ss_ang} ~,
\end{equation}
so that terms $P^{(j)}_{LL'}$ are coefficients in expansion in powers
of $\mu^j$ of contributions of $L,L'$ terms to $P^{s}$.  The $j$ index
has to be even, so the lowest order is either $L+L'$ or $L+L'+1$,
whichever is even.  These terms can be uniquely extracted from
simulations from angular dependence of $P_{LL'}$ terms and so we will
focus on them, although sometimes it is useful to decompose them into
the individual helicity eigenstates instead.  We can write the
redshift-space power spectrum described in terms of $\mu^2$ moments
(equation (\ref{eq:p_ss_ang})) as
\be
P^{s}(\vk) = \sum_{j=0,2,4,\cdots} P_{\mu^j}(k)\mu^{j}. \label{eq:pll_angular_2}
\ee
Note that only even powers of $\mu$ enter in the final expression, as
required by the symmetry.  There is a close relation between the order
of the moments and their angular dependence.  The lowest contribution
in powers of $\mu$ to $P^{s}(k)$ is $P_{00}(\vk)$, which is the only
angular term is isotropic term ($\mu^0$), for $P_{01}(\vk)$ the only
angular term is $\mu^2$, $P_{11}(\vk)$ and $P_{02}(\vk)$ contain both
$\mu^2$ and $\mu^4$ etc. There is always a finite number of terms
contributing to a given order of $j$: for $j=4$ we have 7 terms
contributing to it.

The Legendre multipole expansion is commonly used to analyze the
redshift-space power spectrum in the data analysis.  The motivation
for this expansion is that if one uses full angular information the
Legendre moments are uncorrelated.  Using Legendre polynomials ${\cal
  P}_l(\mu)$, we have
\begin{equation}
  P^{s}(\vk)=\sum_{l=0,2,4,\cdots}P^{s}_l(k){\cal P}_l(\mu) ~, \label{eq:multipole_k1}
\end{equation}
The multipole moments, $P^{s}_l$, are obtained by inversion of this
relation,
\begin{equation}
  P^{s}_l(k)=(2l+1)\int^{1}_{0}P^{s}(\vk){\cal P}_l(\mu)d\mu ~. \label{eq:multipole_k2}
\end{equation}
The errors rapidly grow with the order of the multipole.  The
combination of the monopole ($l=0$), quadrupole ($l=2$), and
hexadecapole ($l=4$) has almost the equivalent cosmological
information to the full 2D spectrum, as argued by \cite{Taruya:2011}.

The difference between the two expansions presented above is just how
to expand the redshift-space power spectrum $P^{s}$, so they are
equivalent if one considers infinite order terms.  They are related to
each other through a simple linear transform, as
\begin{eqnarray}
\left(\begin{array}{c}
P_0^{s}(k) \\ P_2^{s}(k) \\ P_4^{s}(k) \\ P_6^{s}(k) \\  P_8^{s}(k) \\ \vdots
\end{array}\right)
=
\left(\begin{array}{cccccc}
1 & 1/3 & 1/5 & 1/7 & 1/9 & \cdots \\
 0 & 2/3 & 4/7 & 11/21 & 40/99 & \cdots \\
 0 &    0  & 8/35 & 24/77 & 48/143 & \cdots \\
 0 &    0  &   0     &  16/231 &  64/495 & \cdots \\
 0 &	  0 &      0   & 0  & 12/6435 & \cdots \\
 \vdots &  \vdots &  \vdots &  \vdots & \vdots & \ddots
\end{array}\right)
\left(\begin{array}{c}
P_{\mu^0}(k) \\P_{\mu^2}(k) \\P_{\mu^4}(k) \\P_{\mu^6}(k) \\P_{\mu^8}(k) \\ \vdots
\end{array}\right). \label{eq:mu_to_legendre}
\end{eqnarray}
Thus, as addressed in \cite{Seljak:2011}, all terms will contribute to
the monopole $l=0$, all except $P_{00}$ to quadrupole $l=2$, all but
$P_{00}$ and $P_{01}$ to hexadecapole $l=4$, and so on.

%%%%%%%%%%%%%%%%%%%%%%%%%%%%%%%%%%%%%%%%%%%%%%
% Section 3 N-body
%%%%%%%%%%%%%%%%%%%%%%%%%%%%%%%%%%%%%%%%%%%%%%

\section{$N$-body simulations}\label{sec:sim}
\subsection{Dark matter halo and galaxy catalogs}\label{sec:hod}

The power spectra of the derivative expansion are all from
number-weighted velocity moments and thus can be straightforwardly
measured from simulations.  As in \cite{Okumura:2012}, we use a series
of $N$-body simulations of the $\Lambda$CDM cosmology seeded with
Gaussian initial conditions, which is an updated version of
\citep{Desjacques:2009}.  The primordial density field is generated
using the matter transfer function by CMBFAST \cite{Seljak:1996}.  We
adopt the standard $\Lambda$CDM model with the mass density parameter
$\Omega_m=0.279$, the baryon density parameter $\Omega_b=0.0462$, the
Hubble constant $h=0.7$, the spectral index $n_s=0.96$, and a
normalization of the curvature perturbations $\Delta_{\cal
  R}^2=2.21\times 10^{-9}$ (at $k=0.02~{\rm Mpc}^{-1}$) which gives
the density fluctuation amplitude $\sigma_8\approx 0.807$, which are
the best-fit parameters in the WMAP 5-year data
\cite{Komatsu:2009}. We employ $1024^3$ particles of mass
$m_p=2.95\times 10^{11} h^{-1}M_\odot$ in a cubic box of side
$1600\himpc$.  The positions and velocities of all the dark matter
particles are output at $z=0,~0.509,~0.989$, and 2.070, which are
quoted as $z=0,~0.5,~1$, and 2 in what follows for simplicity.  We use
12 independent realizations in order to reduce the statistical
scatters.

Dark matter halos are identified at the four redshifts using the
friends-of-friends algorithm with a linking length equal to 0.17 times
the mean particle separation.  We use all the halos with equal to or
more than 20 particles.  In order to investigate the halo mass
dependences of the clustering measurements, each dark matter halo
catalog is divided into subsamples according to the halo mass, as
$M_{i,{\rm min}}\leq M_i \leq M_{i,{\rm max}}$, where $M_{1,{\rm
    min}}=20\times m_p$ and $M_{i+1,{\rm min}} = M_{i,{\rm
    max}}=3M_{i,{\rm min}}$.  Since the number density of halos is
smaller at higher redshifts, we construct 4 halo subsamples at $z=0$
and 0.5, 3 subsamples at $z=1$, and 2 at $z=2$.

In order to analyze a more realistic sample, we use a halo occupation
distribution (HOD) modeling which populates dark matter halos with
galaxies according to the halo mass \cite[e.g.,][]{Jing:1998a,
  Seljak:2000, Scoccimarro:2001, Berlind:2002, Cooray:2002}.  We
consider a luminous red galaxy (LRG) sample from the Baryon
Oscillation Spectroscopic Survey (BOSS), which is part of Sloan
Digital Sky Survey III (SDSS-III) \cite{Schlegel:2009,
  Eisenstein:2011}.  Galaxies are assigned to the halos using the best
fit HOD parameters for LRGs determined by \cite{White:2011} with the
model of \cite{Zheng:2005}.  The fraction of satellite LRGs populated
in our simulation samples is 12\%, consistent with \cite{White:2011}.
For halos which contain satellite LRGs, we randomly pick up the same
number of dark matter particles to represent the positions and
velocities of the satellites.  As we will see below, the correlation
function measured from the mock LRG catalog is very consistent with
that measured by \cite{White:2011} and \cite{Reid:2012}.  Properties
of the constructed halo and LRG catalogs are summarized in Table
\ref{tab:halo} .

\begin{table}[t!]
\begin{center}
\begin{tabular}{c | ccccccc}
$z$ & Mass & Mass range & $ \bar{N}$ & $\bar{n}$ &$b_1^{mh}$ & $b_1^{hh}$ \\
        & bin & $(10^{12}h^{-1}M_\odot)$ & ($\times10^4$)  & $(h^3{\rm Mpc}^{-3})$ &(cross)&(auto)\\
\noalign{\hrule height 1pt}
0 & $1$ & $5.91 - 17.7$ & 175 & $4.28\times 10^{-4}$ &$1.17$  & 1.19 \\
& $2$ & $17.7-53.2$ & $63.3$ & $1.54\times 10^{-4}$  &$1.46$ & 1.47  \\
&$3$ & $53.2-159$ & $18.7$ & $4.57\times 10^{-5}$& $2.03$ & 1.99  \\
&$4$ & $159-467$ & $4.05$ & $9.89\times 10^{-6}$&$3.04$ & 2.89 \\
\hline
0.5 &$1$ & $5.91 - 17.7$ & $144$ & $3.51\times 10^{-4}$&$1.64$& 1.65 \\
&$2$& $17.7-53.2$ & $44.8$ & $1.09\times 10^{-4}$&$2.16$ & 2.15 \\
&$3$& $53.2-159$ & $9.96$ & $2.43\times 10^{-5}$&$3.12$ & 3.04 \\
&$4$& $159-467$ &  $1.30$ & $3.18\times 10^{-6}$& $4.89$ & 4.72 \\
& LRGs & $5.91-$ \ \ \ \ \ & $125$ & $3.04\times 10^{-4}$& $2.16$ & 2.15  \\
\hline
1 &$1$& $5.91 - 17.7$ &  $101$ & $2.46\times 10^{-4}$& $2.33$ & 2.32 \\
&$2$& $17.7-53.2$ &  $24.9$ & $6.08\times 10^{-5}$& $3.18$ & 3.16 \\
&$3$& $53.2-159$ &  $3.68$ & $8.98\times 10^{-6}$& $4.72$ & 4.70 \\
\hline
2 & $1$& $5.91 - 17.7$ & $25.6$ & $6.25\times 10^{-5}$& 4.65 & 4.73 \\
& $2$ & $17.7-53.2$ & $2.94$ & $7.18\times 10^{-6}$&  6.55 & 6.91 
\end{tabular}
\end{center}
\caption{ Properties of halo catalogs. $\bar{N}$ and $\bar{n}$ are the
  number and number density of halos in each realization,
  respectively. The values $b_1^{mh}$ and $b_1^{hh}$ are the bias
  parameters computed from the cross ($P^{mh}_{00}$) and auto
  ($P^{hh}_{00}$) power spectra, respectively, averaged at $0.01\leq
  k\leq 0.04\hmpci$.  }
\label{tab:halo}
\end{table}

\subsection{Power spectra and two-point correlation functions}

We measure the power spectra of dark matter halos and LRGs from our
simulation samples following \cite{Okumura:2012}.  We assign the
density field and the number-weighted velocity moments in real space
on $1024^3$ grids using a cloud-in-cell interpolation method according
to the positions of particles.  To directly measure $P^{s}_{hh}(\vk)$
we also need the density field in redshift space.  In measuring the
redshift-space density field, we distort the positions of particles
along the line-of-sight according to their peculiar velocities before
we assign them to the grid.  We regard each direction along the three
axes of simulation boxes as the line of sight and the statistics are
averaged over the three projections of all realizations for a total of
36 samples.  We use a fast Fourier transform to measure the Fourier
modes of the density fields in real space $\delta^h(\vk)$ and in
redshift space $\delta_s^h(\vk)$, as well as the number-weighted
velocity moment fields in real space $T_\parallel^{L,h}(\vk)$.  Then
the power spectrum in redshift space, $P^{s}_{hh}(\vk)$, as well as
the power spectra of the number-weighted velocity moments
$P_{LL'}^{hh}(\vk)$, are measured by multiplying the modes of the two
fields (or squaring in case of auto-correlation) and averaging over
the modes within a bin.  To obtain the final estimation of the
auto-power spectrum in real space $P^r_{hh}$ and in redshift space
$P^s_{hh}$, shot noise needs to be subtracted from the measured power
spectrum.  We assume the Poisson model where the contribution of the
shot-noise to the halo power spectrum is described by an inverse of
the halo number density.  Error bars in the following results show the
standard error on the mean.  The dispersion in power spectra
measurements is large on large scales because of sampling variance
\cite{Takahashi:2008}, but it is mostly eliminated by taking the ratio
of the two spectra obtained from the same set of realizations
\cite{McDonald:2009, White:2009, Bernstein:2011}.

Although we mainly analyze the power spectrum in this paper, its
Fourier counter part, the two-point correlation function in redshift
space $\xi_{hh}^s(\vs)$ is also presented for comparison.  We measure
the correlation function using the direct pair counting in
configuration space.

%%%%%%%%%%%%%%%%%%%%%%%%%%%%%%%%%%%%%%%%%%%%%%
% Section 4 Numerical analysis
%%%%%%%%%%%%%%%%%%%%%%%%%%%%%%%%%%%%%%%%%%%%%%

\section{Numerical analysis}\label{sec:analysis}

\subsection{Power spectra of dark matter halos and galaxies}\label{sec:power_spectrum}

\begin{figure}[bt]
\subfigure{\includegraphics[width=1.\textwidth]{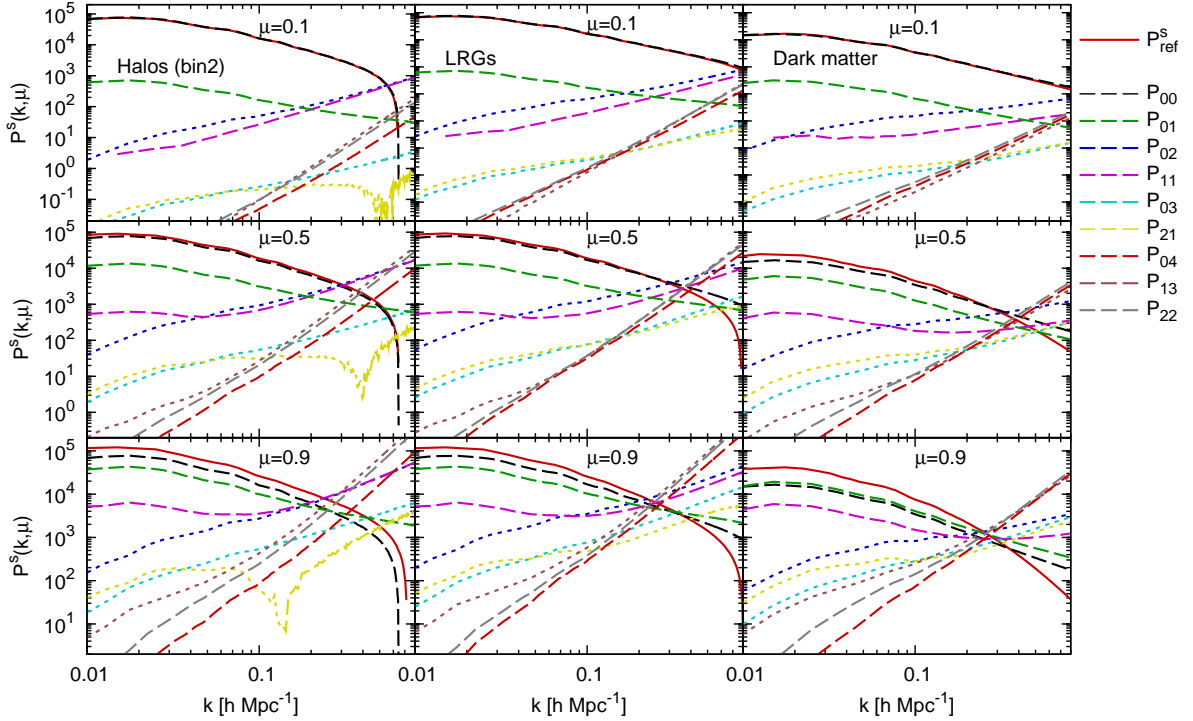}}
\caption{Power spectra measured in redshift space $P^{s}_{\rm ref}(k,\mu)$ and
  individual contributions to it from the terms of the moments
  expansion up to 4-th order at $z=0.5$ for halos (left), LRGs
  (middle) and dark matter (right). The halo subsample has almost the
  same bias value as the LRG sample. The width of $\mu$ bin is 0.2,
  centered around the values shown in each panel.  The dashed lines
  show the positive values while the dotted lines negative values.  }
\label{fig:pkmu_individual_lin}
\end{figure}

We begin by presenting the redshift-space power spectra of halos and
LRGs, $P^{s}_{hh}(\vk)$, directly measured in redshift space.  The
power spectrum in redshift space is shown as functions of $(k,\mu)$ at
$z=0.5$ in figure \ref{fig:pkmu_individual_lin}.  The results are
shown for halos and LRGs at the left and middle columns, respectively.
The mass range of the halo catalog used here is chosen to have
$M_h\sim {2.84}\times 10^{13}h^{-1}M_\odot$ to have the same bias as
LRGs. For comparison, the result for the dark matter, computed in
\cite{Okumura:2012}, is shown in the right column.  In figure
\ref{fig:pkmu_individual_lin} we also show contributions of the terms
of $P_{LL'}$ for $(0\leq L+L' \leq4)$ to $P^{s}(k,\mu)$, computed from
the number density-weighted velocity moments of halos and LRGs.  The
measurement of the angle-averaged power spectrum of momentum,
$P_{11}(k)$, was presented in \cite{Park:2000, Park:2006}.

%%%%%%%%%%%%%%%%%%%%%%%%%%%%%%%%%%%%%%%%%%%%%%
%Figure 1
%%%%%%%%%%%%%%%%%%%%%%%%%%%%%%%%%%%%%%%%%%%%%%

Overall behavior of the power spectra for halos and LRGs is similar to
the dark matter.  At $\mu\sim 0$, contributions from the higher order
power spectra of the velocity moments are small and $P^{s}\simeq
P_{00}$, because each $P_{LL'}$ is multiplied by a factor of
$(k\mu)^{L+L'}$.  On large scales one expects $P_{00}$ to be followed
by the other two linear order terms, which are $P_{01}$ and the scalar
part of $P_{11}$.  The correlators at the same order in powers of
velocity, i.e. equal $L+L'$, contain nontrivial cancellations among
them \cite{Seljak:2011}: higher-order $P_{LL'}$ contain a shot noise
term given by $\bar{n}^{-1}\langle u_\parallel^{L+L'} \rangle$, which
cancels out with other terms of the same order when the total
contribution to the redshift-space power spectrum is considered.  For
example, in figure \ref{fig:pkmu_individual_lin} $P_{11}$ and $P_{02}$
have the similar amplitude but opposite signs, at high $k$.  We do not
see this for dark matter particles which have a much higher number
density than halos and LRGs.

\subsection{Biasing of moments correlators}\label{sec:bias}

For density fluctuations we define bias as the ratio of the power
spectrum of biased objects to that of matter,
$b^2(k)=P^{hh}_{00}(k)/P^{m}_{00}(k)$, with the shot noise subtracted
from the halo spectrum $P^{hh}_{00}$.  Following \cite{Seljak:2011},
we can generalize the concept of the bias to
\be
b_{LL'}^{hh}(\vk)=\frac{P^{hh}_{LL'}(\vk)}{P^{mm}_{LL'}(\vk)}. \label{eq:bias}
\ee
On sufficiently large scales where linear theory is believed to be
applicable, we have $b_{00}^{hh}=b_1^2$, $b_{01}^{hh}=b_1$ and
$b_{11}^{hh}=1$, independent of scale and angle.  All of the bias
terms can be alternatively defined using the cross power spectrum, for
example we can define $b_{00}^{mh}$ as
\be
(b_{00}^{mh}(k))^{1/2}=\frac{P^{mh}_{00}(k)}{P^{mm}_{00}(k)}, \label{eq:b_mh}
\ee
which is free from the shot noise issues. At large scales where the
linear theory holds, $b_{00}^{hh}=b_{00}^{mh}$.Since we do not expect
shot noise to be an issue for higher order moments correlators (as
discussed in more detail below) we only look at the density-density
correlations using the cross-correlations with the dark matter.
 
 %%%%%%%%%%%%%%%%%%%%%%%%%%%%%%%%%%%%%%%%%%%%%%
%Figure 2
\begin{figure}%[!t]
\subfigure{\includegraphics[width=1.\textwidth]{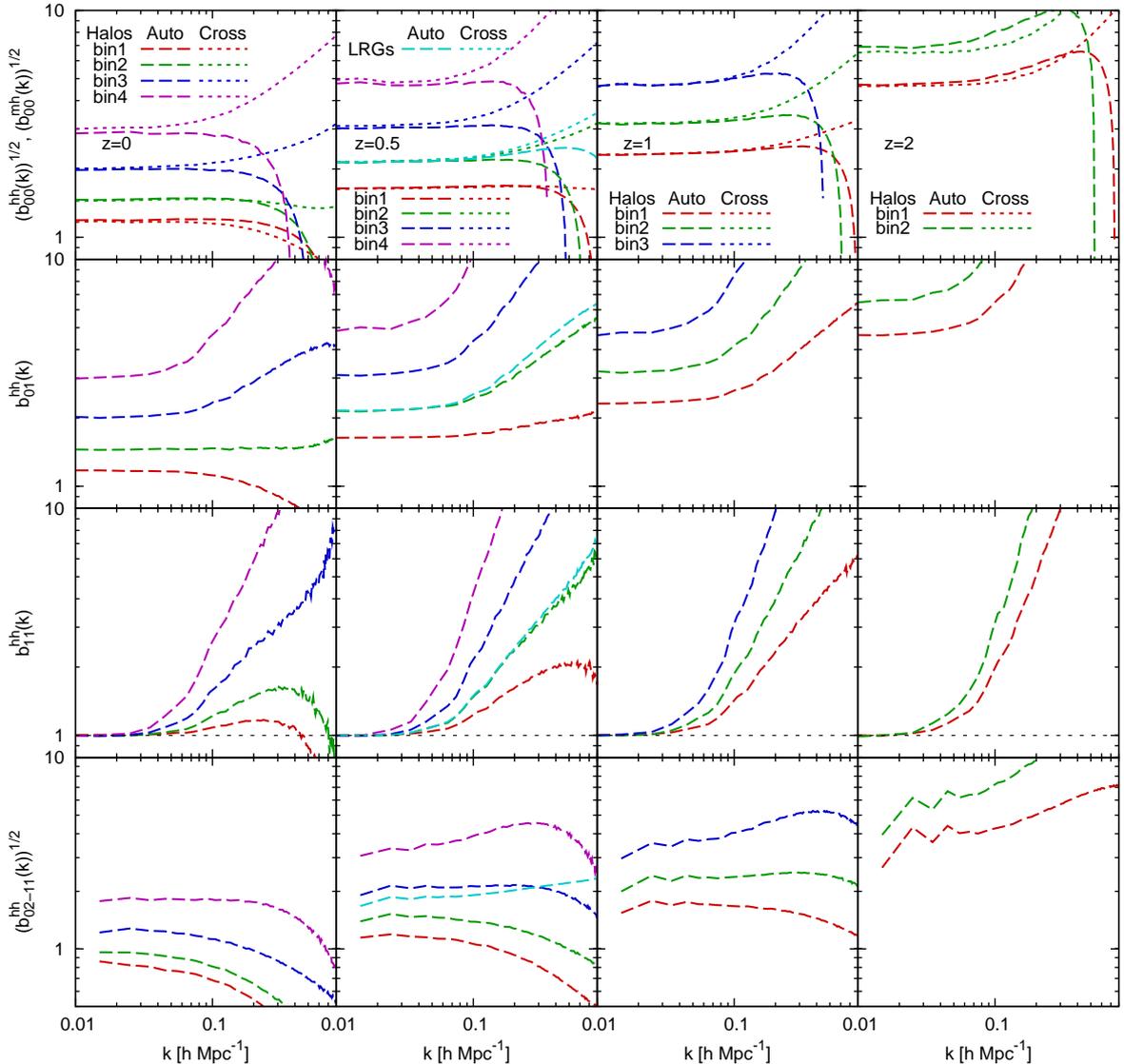}}
\caption{Bias parameters, $(b^{hh}_{00})^{1/2}(k)$ and
  $(b^{mh}_{00})^{1/2}(k)$ (top), $b^{hh}_{01}(k)$ (second),
  $b^{hh}_{11}(k)$ (third), and $b^{hh}_{02-11}(k)$ (bottom) for halos
  and LRGs.  The light blue lines at $z=0.5$ show the results for
  LRGs, while all the other lines for halos of different mass bins.
  In the top panels the bias parameters computed using the auto
  ($P^h_{00}$) and cross ($P^{mh}_{00}$) power spectra are plotted as
  the dashed and dotted lines, respectively.  }
\label{fig:bias}
\end{figure}
%%%%%%%%%%%%%%%%%%%%%%%%%%%%%%%%%%%%%%%%%%%%%%

Here we wish to investigate the scale dependence of these generalized
bias parameters.  Figure \ref{fig:bias} shows the halo bias defined
above and determined from simulations.  The top panels show
$(b_{00}^{hh})^{1/2}(k)$, shot noise corrected (eq. (\ref{eq:bias})
with $L=L'=0$), and $(b_{00}^{mh}(k))^{1/2}$, the bias from the
halo-matter cross-correlation (eq. (\ref{eq:b_mh})).  One can see that
the two bias parameters agree on large scales and for low halo mass,
corresponding to high halo number density.  The more massive halos are
more sparse, the shot noise correction becomes larger, and there is a
difference between the two halo bias estimates even on very large
scales.  In addition, the bias estimated from the auto-correlation is
strongly suppressed on small scales due to the shot noise correction
(e.g. \cite{Smith:2007}).  We assume the bias to be constant on
sufficiently large scales and determine the large-scale bias
$b_1^{mh}$ and $b_1^{hh}$ averaging the data points over $0.01\leq
k\leq 0.04\hmpci$, shown in Table \ref{tab:halo}. We will use bias
from cross-correlation as the true bias, assuming that the bias from
auto-correlation suffers from imperfect shot noise subtraction.

Let us consider the next-order bias, $b_{01}^{hh}$, defined as
\be
b_{01}^{hh}(k)=\frac{P_{01}^{hh}(\vk)}{P_{01}^{mm}(\vk)} = \frac{P_{01}^{(2)hh}(k)}{P_{01}^{(2)mm}(k)} .
\ee
Although $b_{LL'}^{hh}$ has angular dependences by definition in
general, $b_{01}^{hh}$ is a function of $k$ only as is well known for
$b_{00}^{hh}$ and $b_{00}^{mh}$.  In the second panels of figure
\ref{fig:bias}, the bias parameter, $b_{01}^{hh}(k)$, is shown.  In
linear theory $b_{01}^{hh}=(b_{00}^{mh})^{1/2}=b_1$, so
$b_{01}^{hh}=(b_{00}^{mh})^{1/2}$ with a percent level accuracy on
sufficiently large scales, as expected.  However, we also see a strong
scale dependence of this bias, which is worse for the more biased
halos.

For the next-order bias, $b_{11}^{hh}$, we need to separate the
$\mu^2$ and $\mu^4$ parts.  As detailed in \cite{Seljak:2011}, the
$\mu^2$ part originates from the auto-correlation of the vector part
of the momentum density, $P_{11}^{(2)}=P^{1,1,1}_{1,1}$. It is a
nonlinear term, since vector component of momentum density is zero in
linear theory.  The $\mu^4$ term contains a linear order contribution,
which is often described as the velocity auto-correlation.  Thus we
define $b_{11}^{hh}$ using the anisotropic term of $P_{11}$ as
\be
b_{11}^{hh}(k)= \frac{P^{(4)hh}_{11}(k)}{P^{(4)mm}_{11}(k)} .
\ee
The bias defined in this way has no angular dependence.  The third
panels of figure \ref{fig:bias} show the bias $b_{11}^{hh}$.  We see
that $b_{11}^{hh}$ starts to deviate from linear theory predictions at
lower $k$ than $b_{01}^{hh}$ or $b_{00}^{mh}$ and these deviations are
more important for more massive halos. These effects are large: for
moderately biased halos with $b=2$ they are nearly a factor of 2 at
$k=0.1\hmpci$ at $z=0$. It is obvious that halos do not measure
velocity-velocity correlations except on very large scales.

At the next order we have $P_{02}$, which contributes to the $\mu^2$
term ($P_{02}^{(2)}$) and the $\mu^4$ term ($P_{02}^{(4)}$) in
$P^{s}$.  The more important $\mu^2$ term contains a shot noise
contribution $\sigma_v^2/n$, where $\sigma_v^2$ is the velocity
dispersion.  $P_{11}^{(2)}$ also contains the same shot noise term
which cancels out that of $P_{02}^{(2)}$, as discussed in section
\ref{sec:power_spectrum}.  Neither of these terms, nor any of the
higher order terms, contain any linear order contributions, so we do
not expect them to contain cosmologically useful information, but it
is important to understand them to estimate the nonlinear effects.  We
will thus combine the two $\mu^2$ terms into a bias term
\be
b_{02-11}^{hh}(k)=
\frac{P^{(2)hh}_{02}(k)-P^{(2)hh}_{11}(k)}{P^{(2)mm}_{02}(k)-P^{(2)mm}_{11}(k)}.
\ee
This is also shown in figure \ref{fig:bias}. We see that this bias
scales strongly with density bias $b_{00}^h \sim b_1^2$.

%%%%%%%%%%%%%%%%%%%%%%%%%%%%%%%%%%%%%%%%%%%%%%
%Figure 3
\begin{figure}%[!t]
\subfigure{\includegraphics[width=1.\textwidth]{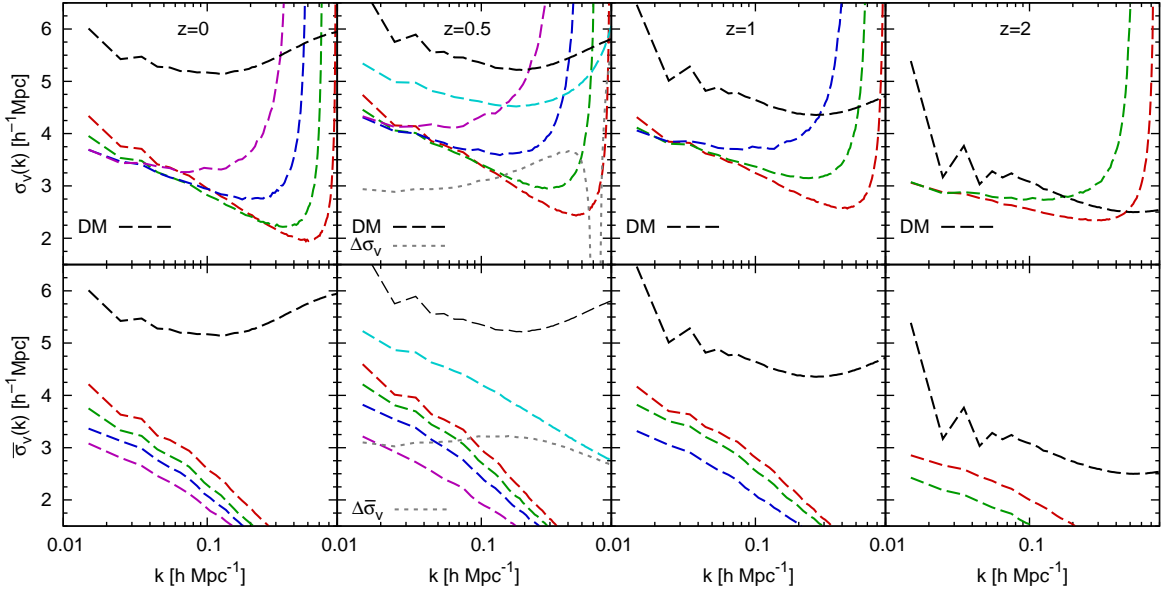}}
\caption{Velocity dispersion parameter with (upper panels) and without
  (lower panels) shot noise.  The color of each line corresponds to
  the one with the same color in figure \ref{fig:bias}.  The black
  line is for dark matter.  The quantity $\Delta\sigma_v$ and
  $\Delta\bar{\sigma}_v$ are respectively shown in the top and bottom
  panels for $z=0.5$ as the dotted gray line (see the text).  }
\label{fig:sigmav}
\end{figure}
%%%%%%%%%%%%%%%%%%%%%%%%%%%%%%%%%%%%%%%%%%%%%%

Fingers-of-god (FoG) are caused by small scale velocity dispersion,
which smears the galaxies along the line of sight and lead to a
suppression of the linear terms in RSD power spectrum. The
phenomenological model often used is that the linear terms, $P_{00}$,
$P_{01}$ and $P_{11}$, are multiplied by FoG kernel which is often
modeled as a Gaussian or a Lorentzian, and which is a function of
$k^2\mu^2\sigma_v^2$, where $\sigma_v$ is the velocity dispersion.
This velocity dispersion is sometimes modeled using linear theory
averaged velocity squared $\langle v^2 \rangle=\int P(k)dk/2\pi^2$,
but as we show below this quantity is not relevant.  The first term
that is suppressed is $P_{00}$ and the leading order FoG term has a
$k^2\mu^2$ dependence, $-k^2\mu^2\sigma_v^2P_{00}$. Since the only two
terms besides $P_{01}$ leading to $\mu^2$ dependence in $P^s(k,\mu)$
are $P_{11}$ and $P_{02}$, and since these two terms each contain a
shot noise that is canceled out in their total contribution to
$P^s(k,\mu)$, it is natural to define the velocity dispersion as
\be
\sigma_v(k) = 
\sqrt{ \frac{P^{(2)hh}_{02}(k)-P^{(2)hh}_{11}(k)}{P^{hh}_{00}(k)} }.
   \label{eq:sigmav}
\ee
The same quantity but with the shot noise included to the denominator
$P^{hh}_{00}$ is denoted as $\bar{\sigma}_v(k)$.  We show the velocity
dispersion parameter of halos and LRGs with the shot noise subtracted
from and included to $P^{hh}_{00}$ in the top and bottom of figure
\ref{fig:sigmav}, respectively.  For comparison, the velocity
dispersion parameter computed for dark matter is also plotted at the
both panels of the figure.

First thing to notice is that the velocity dispersion of halos is
significantly smaller than that of dark matter or LRGs. This is to be
expected, since halos have no small scale nonlinear velocity
dispersion, as they are defined as the center of mass of the dark
matter particles inside the halo. In contrast, LRG satellites and dark
matter particles both stream around the halo center with high virial
velocities inside the halos, which leads to a significant velocity
dispersion.  The velocity dispersion is nearly independent of the halo
bias, as expected in the model above.

Second thing to notice is that velocity dispersion of halos is rapidly
decreasing towards smaller scales and has no relationship to $\langle
v^2 \rangle=\int P(k)dk/2\pi^2 \sim (6Mpc/h)^2$ for this model at
$z=0$.  Physically this makes sense: any large scale bulk motions that
contribute to $\langle v^2 \rangle$ should have no effect on the
relative velocity dispersion between two close particles, since they
move both particles by the nearly same velocity. Thus there must be a
significant reduction of velocity dispersion and FoG from bulk
velocities as we go to higher $k$ and only the really small scale
velocity dispersion from inside the halos can contribute to FoG on
smaller scales. Even that small scale velocity dispersion must cancel
out if one considers very high $k$, which is dominated by the zero lag
correlations, i.e. the shot noise term.  The exact cancellation
between the shot noise of $P_{02}$ and $P_{11}$ is thus required
\cite{Seljak:2011}.  This also shows that we must include $P_{11}$ in
the FoG definition, since it partially cancels out $P_{02}$.

We can try to define the true small scale velocity dispersion of LRGs
by using the halo sample which has the same bias as the LRGs at large
scales, defining a quantity $\Delta \sigma_v$ as
\be
\Delta\sigma_v = \sqrt{\sigma_{v, {\rm LRG}}^2 - \sigma_{v,h}^2}, \label{eq:deltasigma}
\ee
where $\sigma_{v,{\rm LRG}}$ is the velocity dispersion for LRGs and
$\sigma_{v,h}$ is that for halos with the same bias as LRGs.  The same
quantity which includes the shot noise to $P^{hh}_{00}$, $\Delta
\bar{\sigma}_v$, can be defined as well.  The resulting $\Delta
\sigma_v$ and $\Delta \bar{\sigma}_v$ are respectively shown at the
top and bottom of figure \ref{fig:sigmav} for $z=0.5$ as the dotted
gray line.  The difference of the velocity dispersion between LRGs and
halos is nearly constant at $k<0.1\hmpci$: this is expected, since we
argued that small scale velocity dispersion does not get canceled as
quickly as the bulk motions. At higher $k$ even the small scale
velocity dispersion cancels out.

%%%%%%%%%%%%%%%%%%%%%%%%%%%%%%%%%%%%%%%%%%%%%%
%Figure 4
\begin{figure}%[!t]
\subfigure{\includegraphics[width=1.\textwidth]{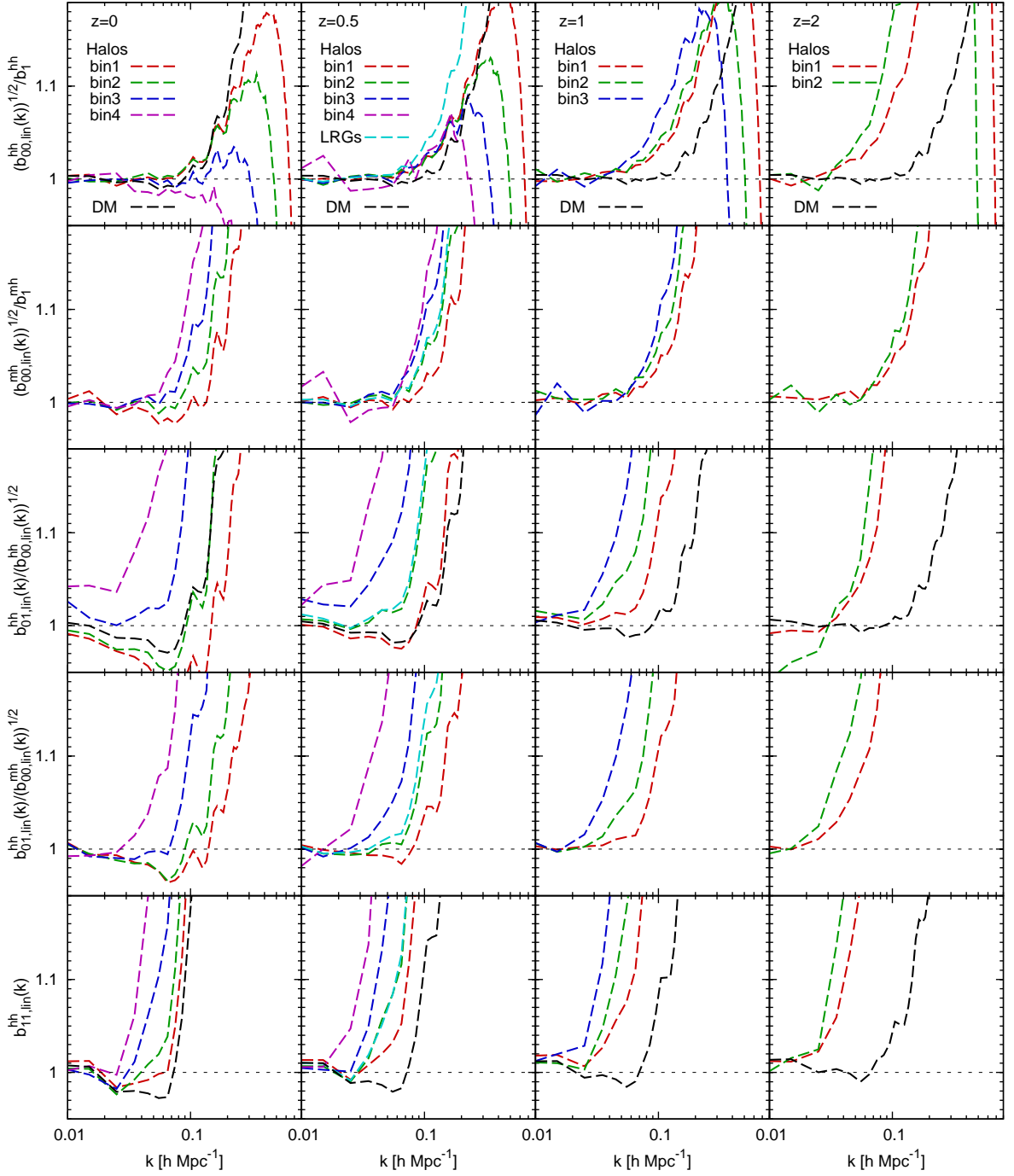}}
\caption{Same as figure \ref{fig:bias}, but the three power spectra
  which contain linear order contributions, $P^{mm}_{00}$,
  $P^{mm}_{01}$ and $P^{mm}_{11}$ in the definitions of bias
  parameters, are replaced by the linear theory power spectrum: from
  the top to bottom row, $(b^{hh}_{00,{\rm lin}})^{1/2}(k)/b_1^{hh}$,
  $(b^{mh}_{00,{\rm lin}})^{1/2}(k)/b_1^{mh}$, $b^{hh}_{01,{\rm
      lin}}(k)/(b^{mh}_{00})^{1/2}$, $b^{hh}_{01,{\rm
      lin}}(k)/(b^{hh}_{00})^{1/2}$, and $b^{hh}_{11,{\rm lin}}(k)$.
  The horizontal lines at the value of unity in each panel show the
  prediction for these quantities from linear theory with the input
  cosmological parameters in our simulations.  }
\label{fig:bias_linear}
\end{figure}
%%%%%%%%%%%%%%%%%%%%%%%%%%%%%%%%%%%%%%%%%%%%%%

\subsection{Relation to linear RSD}\label{sec:bias_lin}

In previous subsection we defined the bias of the velocity moment
correlators relative to the dark matter.  In order to investigate the
ability of RSD to recover linear theory predictions we can also define
the generalized halo bias alternatively for $(LL')=(0,0)$, $(0,1)$ and
$(1,1)$ as
\be
b_{LL',{\rm lin}}^{hh}(\vk)=\frac{P^{hh}_{LL'}(\vk)}{P^{mm}_{LL', {\rm lin}}(\vk)} \label{eq:bias_lin},
\ee
and the bias using the matter-halo cross power spectrum as
\be
(b_{00}^{mh}(k))^{1/2} = \frac{P^{mh}_{00}(k)}{P^{mm}_{00, {\rm lin}}(k)},
\ee
where $P^{mm}_{00, {\rm lin}}(k)$ is simply the linear mass power
spectrum, and its relation to $P^{mm}_{01}$ and $P^{mm}_{11}$ is given
in equation (\ref{eq:pll_kaiser}).  These bias parameters based on the
linear power spectrum for dark matter are shown in figure
\ref{fig:bias_linear}.  We also show the nonlinear dark matter power
spectrum relative to its linear spectrum as the black dotted curve.
It is an auto correlation of dark matter thus defined as
$b^{mm}_{LL',{\rm lin}}(k)=P^{mm}_{LL'}(k)/P^{mm}_{LL',{\rm lin}}(k)$.
Note that unlike the bias parameters $b^{hh}_{LL'}$ and $b^{mh}_{LL'}$
presented in section \ref{sec:bias}, the parameters $b^{hh}_{LL',{\rm
    lin}}$ and $b^{mh}_{LL',{\rm lin}}$ suffer from sampling variance
at large scales.  In order to reduce the large scatter because of
sampling variance at $k\leq 0.04\hmpci$, we divide them by
$b^{mm}_{LL'{\rm lin}}(k)$ measured at redshift $z=8.5$ where the
density perturbation is known to be linear at such scales.

The top row of figure \ref{fig:bias_linear} show the density bias
parameter, $(b_{00,{\rm lin}}^{hh})^{1/2}$ as well as the bias for
dark matter, $(b_{00,{\rm lin}}^{mm})^{1/2}$, while in the second row
we show $(b_{00,{\rm lin}}^{mh})^{1/2}$.  As is clear from the plots
of $(b_{00,{\rm lin}}^{mm})^{1/2}$, the nonlinearity of $P^{mm}_{00}$
starts to be significant at $k\simeq 0.1\hmpci$ for $z=0$ and $k\simeq
0.2\hmpci$ for $z=2$.

In third row of figure \ref{fig:bias_linear} we compare the normalized
first-order bias $b^{hh}_{01,{\rm lin}}$ to $(b_{00,{\rm
    lin}}^{hh})^{1/2}$ and in fourth row to $(b_{00,{\rm
    lin}}^{mh})^{1/2}$, where
\be
\frac{b^{hh}_{01,{\rm lin}}(k)}{ (b^{mh}_{00,{\rm lin}}(k))^{1/2} } = \frac{kP_{01}^{(2)hh}(k)}{{\cal H}fP_{00}^{mh}(k)}.
\label{eq:b01_lin_h}
\ee
We see that for very low $k$ all the quantities, including the dark
matter, are equal in the fourth row.  This means that $P_{01}$ is
tracing the true large scale halo bias as defined by the halo-matter
cross-correlation.  In contrast, the values in the third row differ at
a level of a few percent, suggesting that the halo auto-power spectrum
with the standard shot noise subtraction does not trace the true halo
bias, most likely because the shot noise is not $\bar{n}^{-1}$, where
$\bar{n}$ is the halo density. Since auto-correlation is measurable
while cross-correlation with the dark matter is not (unless we have
weak lensing measurements) this issue needs to be taken into account
when analyzing real surveys. For LRGs the effect is below 1\%.  In
both rows we see that there is strong scale dependence of the bias:
for LRGs it is 10\% at $k=0.1\hmpci$.

The next-order quantity relevant to the growth rate measurement is, 
\be
b^{hh}_{11,{\rm lin}}(k)=\frac{k^2P_{11}^{(4)hh}(k)}{{\cal H}^2f^2 P_{00,{\rm lin}}^{mm}(k)}.
\ee
This term has $\mu^4$ angular dependence and dominates the
hexadecapole.  Although such a higher-order statistics is generally
noisier than the lower-order statistics above, measuring $P_{11}$ has
a potential to enable us to directly give constraints on $f$ thus on
modified gravity models, independently from the galaxy biasing or the
existence of the shot noise.  We show the resulting parameter
$b^{hh}_{11,{\rm lin}}$ at the bottom of figure \ref{fig:bias_linear}.
At very low $k$ it approaches the true value, but the scale dependence
is stronger than for $b_{01,{\rm lin}}^{hh}$: only the lowest $k$
modes trace the dark matter velocity. For LRGs the nonlinear effect is
10\% at $k=0.06\hmpci$.

%%%%%%%%%%%%%%%%%%%%%%%%%%%%%%%%%%%%%%%%%%%%%%
%Figure 5
\begin{figure}%[bt]
\centering
%\resizebox{\hsize}{!}{
\subfigure{\includegraphics[width=1.\textwidth]{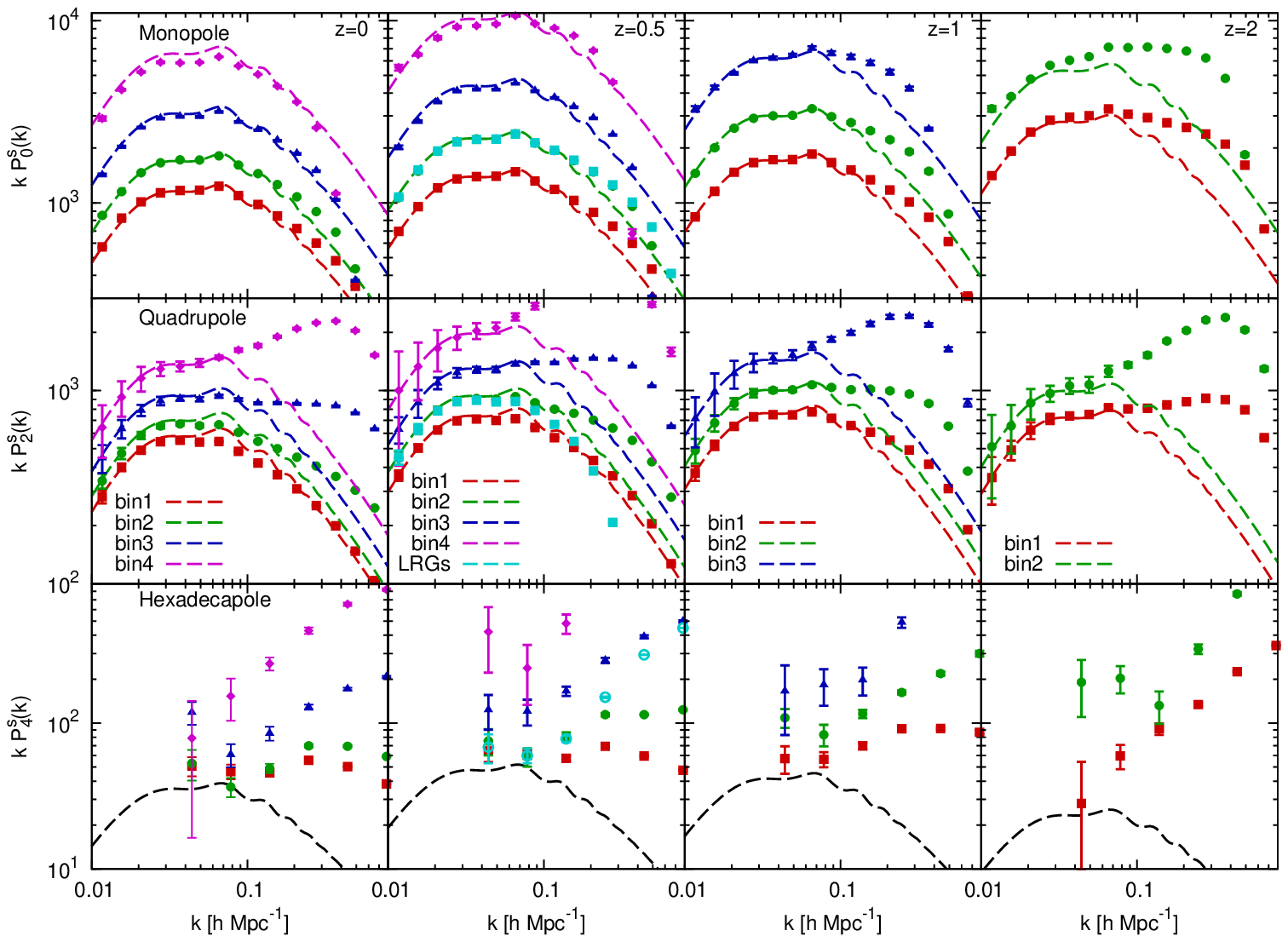}}
\subfigure{\includegraphics[width=1.\textwidth]{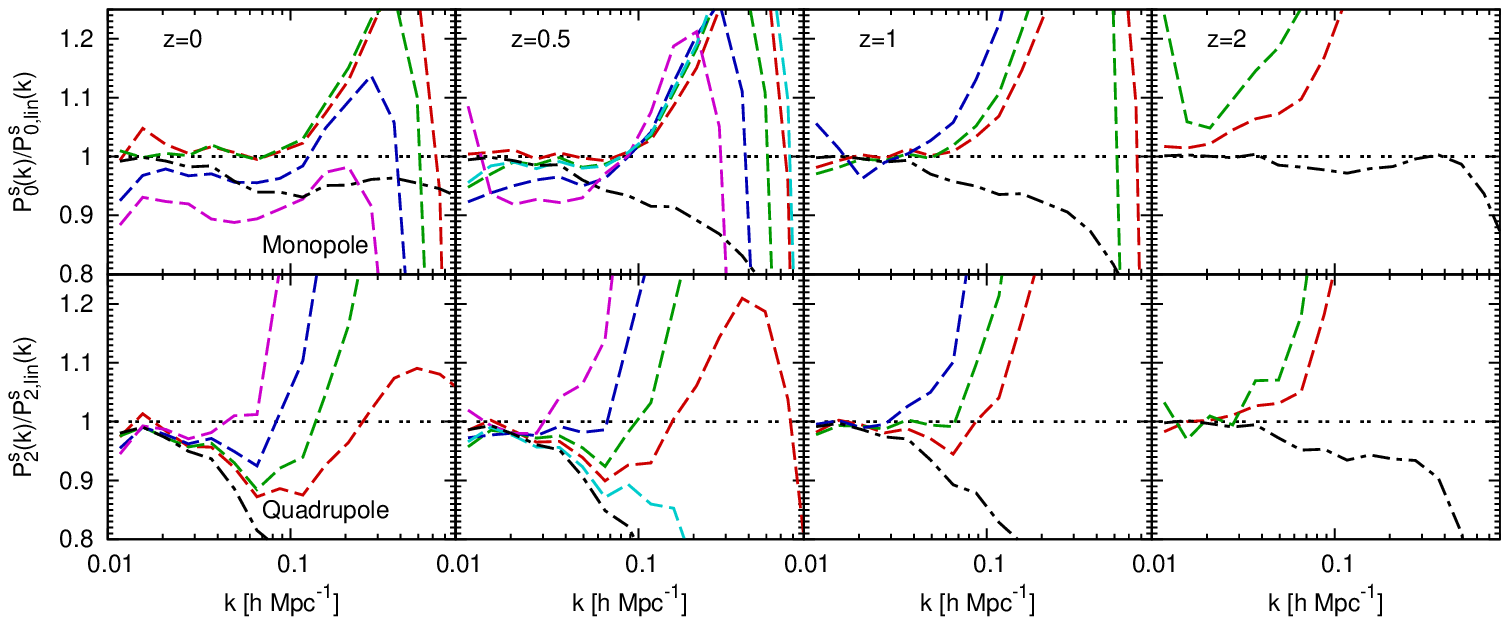}}
\caption{{\it Top set:} Multipole power spectra of halos and LRGs.
  The corresponding linear theory predictions are shown as the dashed
  line with the same color for monopoles and quadrupoles, while as the
  black line for hexadecapoles.  Artificial cuts are put for the plots
  of the hexadecapoles at low $k$ because of large sampling variance.
  {\it Bottom set:} Monopole and quadrupole spectra divided by linear
  theory.  The color of the lines corresponds to the one with the same
  color in the top set.  The results for dark matter, obtained in
  \cite{Okumura:2012}, are also shown as the dot-dashed black lines
  for comparison.  }
\label{fig:multipole}
\end{figure}
%%%%%%%%%%%%%%%%%%%%%%%%%%%%%%%%%%%%%%%%%%%%%%

\subsection{Legendre moments of redshift-space power spectrum}\label{sec:legendre}

In this subsection we present the power spectrum directly measured in
redshift space in section \ref{sec:power_spectrum}, $P^s(k,\mu)$, in
terms of the Legendre multipole spectra.  In the top set of figure
\ref{fig:multipole} we show the resulting multipole spectra of
mass-binned halos and LRGs. The top, middle and bottom panels
respectively show the monopole, quadrupole and hexadecapole spectra.
Monopole is shot noise subtracted using the standard shot noise
$n^{-1}$.  We show the corresponding linear theory predictions,
$P^s_{l,{\rm lin}}=b_1P^{s,m}_{l,{\rm lin}}$.  For the bias $b_1$ we
use the values of $(b^{mh}_{00}(k))^{1/2}$ (equation (\ref{eq:b_mh}))
at large scales determined in section \ref{sec:bias} and quoted in
table \ref{tab:halo}.  The ratio to linear theory is shown in the
bottom of figure \ref{fig:multipole}.  As in the previous subsection,
the effect of sampling variance on the ratio is prominent.  We thus
follow the same process: we divide the ratio of multipoles to linear
theory by that computed at $z=8.5$ for the scales $k\leq 0.04\hmpci$.
Once again we see deviations from linear theory in the monopole on
very large scales as a consequence of inaccurate shot noise
subtraction in the auto-correlation.  The halo monopole is typically
first above the linear theory predicts due to the nonlinear and halo
biasing effects, before it comes down because of the shot noise
subtraction.  In contrast, the dark matter monopole as measured in
\cite{Okumura:2012} is flat or decreased towards the smaller scales,
shown at the bottom two rows of figure \ref{fig:multipole} as the
dot-dashed lines.

The quadrupole moments show a similar behavior.  They are not affected
by the shot noise so there is no uncertainty with the exact value of
the shot noise.  As a result all of the halo quadrupole moments agree
on vary large scales, similar to the situation in $P_{01}$. We note
that because the quadrupole moment integrates over all the modes with
a non-positive function ${\cal P}_2(\mu)=(3\mu^2-1)/2$, the result is a partial
cancellation of modes and the quadrupole is more susceptible to the
sampling variance: these effects reach 10\% at $k \sim 0.02\hmpci$.
However, by dividing it by the quadrupole for dark matter at very high
redshift, such sampling variance effects are well eliminated as shown
on our plots.  The quadrupole deviates from the linear theory
prediction at larger scales than that for the monopole and typically
the biasing and nonlinear effects make the halo quadrupole increase
relative to the linear theory. In contrast, for the LRG sample and for
the dark matter the effect is a strong suppression of quadrupole
relative to the linear theory prediction.  At the largest scales, the
all the quadrupole spectra approach linear theory at $z=1$ and 2, and
just a few percent below at $z=0$ and $z=0.5$.  We also show in figure
\ref{fig:multipole} the hexadecapole. It is even noisier than the
quadrupole \cite{Nishimichi:2011} so we adopt $k$ binning for it twice
as large as for the lower multipoles and put artificial cuts for the
plot for $k<0.04\hmpci$.

%%%%%%%%%%%%%%%%%%%%%%%%%%%%%%%%%%%%%%%%%%%%%%
% Section 5 Expansion in power of mu^2
%%%%%%%%%%%%%%%%%%%%%%%%%%%%%%%%%%%%%%%%%%%%%%

\section{Expansion in powers of $\mu^2$} \label{sec:mu}
In previous work \cite{Okumura:2012}, using dark matter simulations,
we have seen that the series expansion of equation \ref{eq:p_ss_halo}
is convergent for $k\mu\sigma_v/{\cal H} < 1$ but there is no
convergence for $k\mu\sigma_v/{\cal H} > 1$, where $\sigma_v$ is a
typical rms velocity.  This series expansion could be more convergent
for halos than for dark matter since, as we have established, halo
centers are not sensitive to the the velocity dispersion inside halos.
The convergence is an issue if we want to investigate $P^s(k,\mu)$ or
its Legendre moments. Alternatively, we can consider an expansion in
powers of $\mu^2$, which enables us to sidestep the issue of
divergence of the terms \cite{Seljak:2011}: for any finite power of
$\mu^2$ there is a finite number of $P_{LL'}$ terms contributing to
it.  In \cite{Okumura:2012} we tested the powers of $\mu^2$ expansion
against the dark matter simulations.  Here we want to apply this to
halos and galaxies.

As is clear from equation (\ref{eq:kaiser}), only the three lowest
terms, $\mu^0$, $\mu^2$, and $\mu^4$, contain cosmological information
at the linear order, so in principle these are the only relevant
terms.  However, if we expand the full $P^{s}(k,\mu)$ into powers of
$\mu^2$ and try to determine the coefficients from the data, the
resulting coefficients will be correlated: only Legendre expansion
assures uncorrelated values. As a result there will be mixing of
higher powers of $\mu^2$ into lower powers if they are not accounted
for in the fits, or there will be strong degeneracies and the fits
will be unstable if all the coefficients are accounted for but we
allow them to take any value.  At high $k$ it may be more advantageous
to resum the higher $\mu^2$ terms into the so called FoG kernel
\cite{Okumura:2012} and fit for that instead.

%%%%%%%%%%%%%%%%%%%%%%%%%%%%%%%%%%%%%%%%%%%%%%
%Figure 6
\begin{figure}%[!t]
\centering
\subfigure{\includegraphics[width=1.\textwidth]{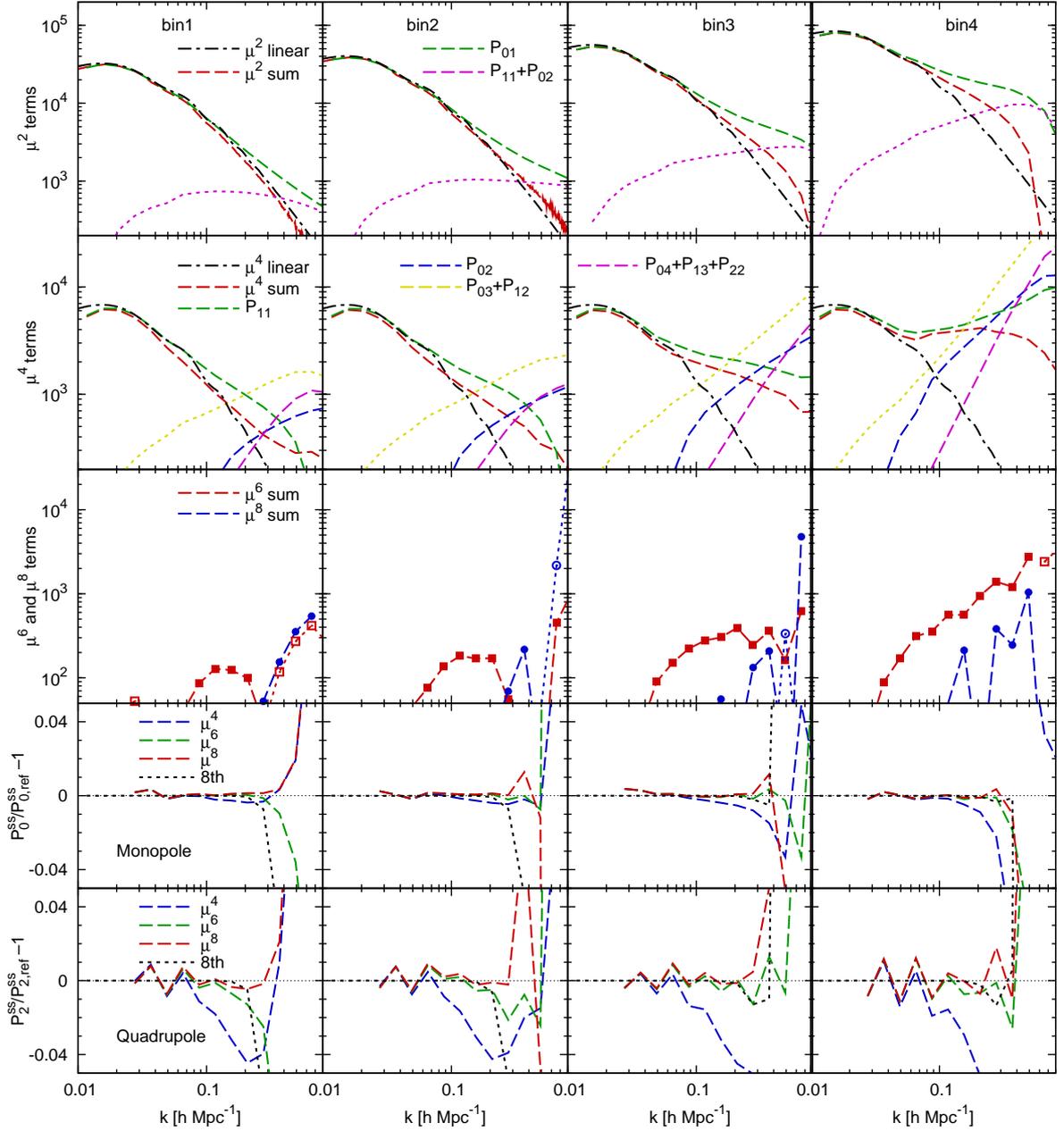}}
\caption{Contributions to $\mu^2$ (top), $\mu^4$ (second), and $\mu^6$
  and $\mu^8$ (third) terms in redshift-space power spectrum
  $P^{s}(k,\mu)$ at $z=0$.  The positive and negative contributions
  are shown as dashed and dotted lines, respectively, and additionally
  as the filled and open points for $\mu^6$ and $\mu^8$ terms.  Linear
  theory predictions for $\mu^2$ and $\mu^4$ terms are shown as the
  dot-dashed curves.  The bottom two rows show the error in the
  monopole and quadrupole as a function of order in powers of $\mu^2$:
  up to $\mu^4$ (blue), $\mu^6$ (green) and $\mu^8$ (red) terms using
  equation (\ref{eq:mu_to_legendre}). Because of large sampling variance, 
  we put artificial cuts at large scales. Error between the summed power
  spectrum up to 8-th order and the reference spectrum is also shown
  as the dotted lines.  }
\label{fig:mu_z007}
\end{figure}
%%%%%%%%%%%%%%%%%%%%%%%%%%%%%%%%%%%%%%%%%%%%%%

%%%%%%%%%%%%%%%%%%%%%%%%%%%%%%%%%%%%%%%%%%%%%%
%Figure 7
\begin{figure}%[!t]
\centering
\subfigure{\includegraphics[width=1.\textwidth]{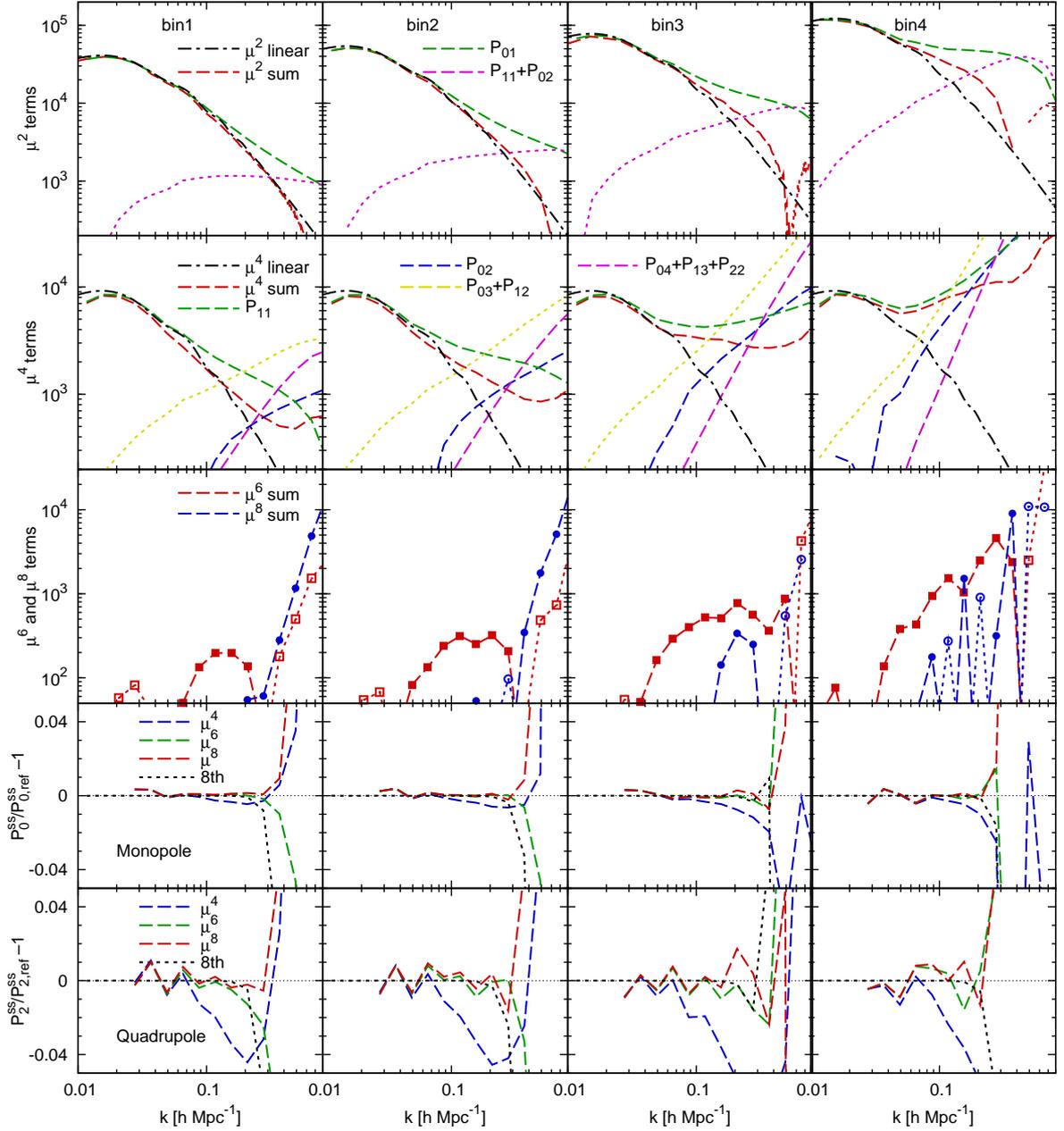}}
\caption{Same as figure \ref{fig:mu_z007} but for $z=0.5$. The results
  for LRGs are separately shown in figure \ref{fig:mu_lrg}.  }
\label{fig:mu_z005}
\end{figure}
%%%%%%%%%%%%%%%%%%%%%%%%%%%%%%%%%%%%%%%%%%%%%%

%%%%%%%%%%%%%%%%%%%%%%%%%%%%%%%%%%%%%%%%%%%%%%
%Figure 8
\begin{figure}%[!t]
\centering
\subfigure{\includegraphics[width=1.\textwidth]{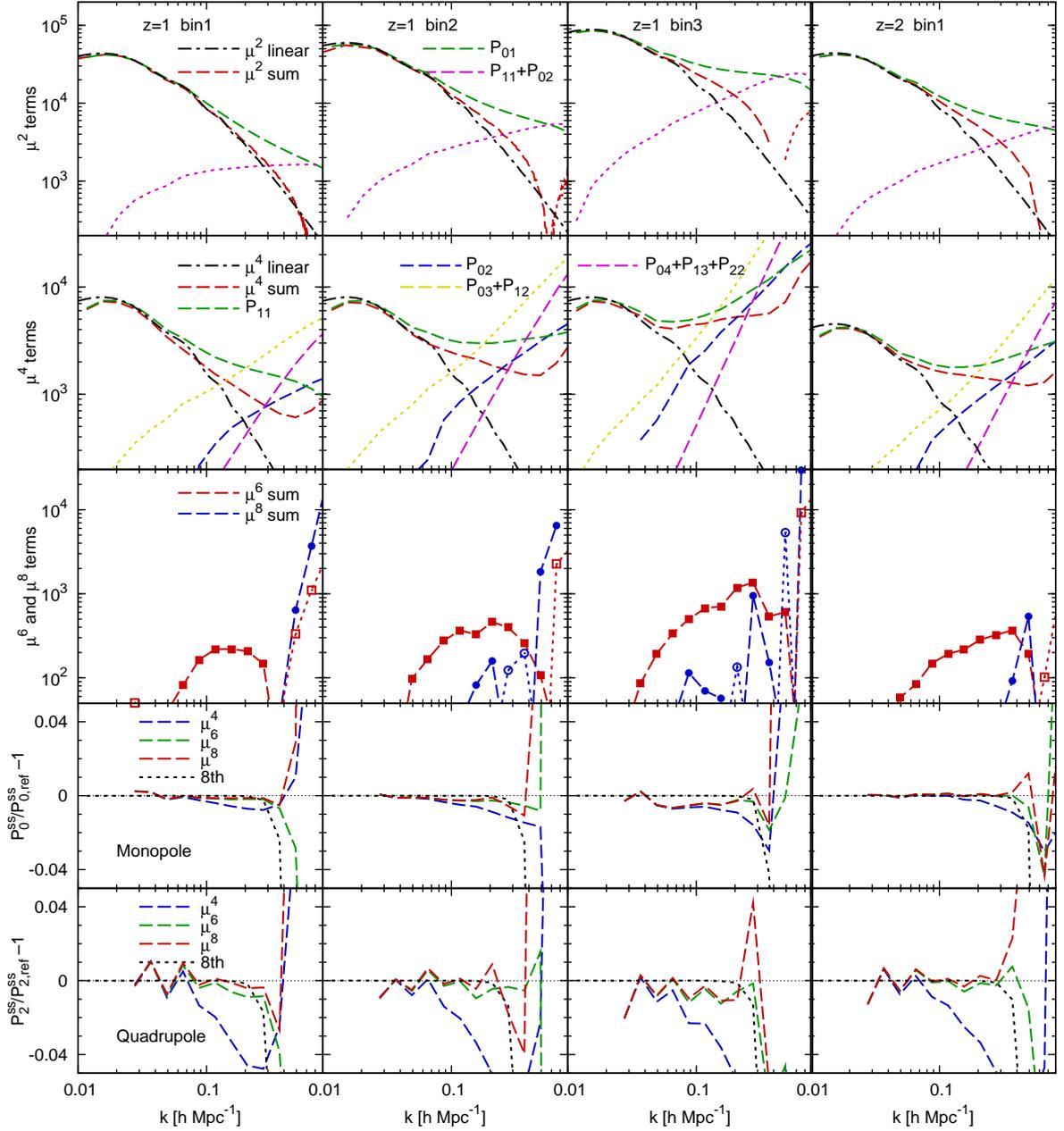}}
\caption{Same as figure \ref{fig:mu_z007} but for $z=1$ and $z=2$. 
}
\label{fig:mu_z004}
\end{figure}
%%%%%%%%%%%%%%%%%%%%%%%%%%%%%%%%%%%%%%%%%%%%%%

%%%%%%%%%%%%%%%%%%%%%%%%%%%%%%%%%%%%%%%%%%%%%%
%Figure 9
\begin{figure}%[!t]
\centering
\subfigure{\includegraphics[width=1.\textwidth]{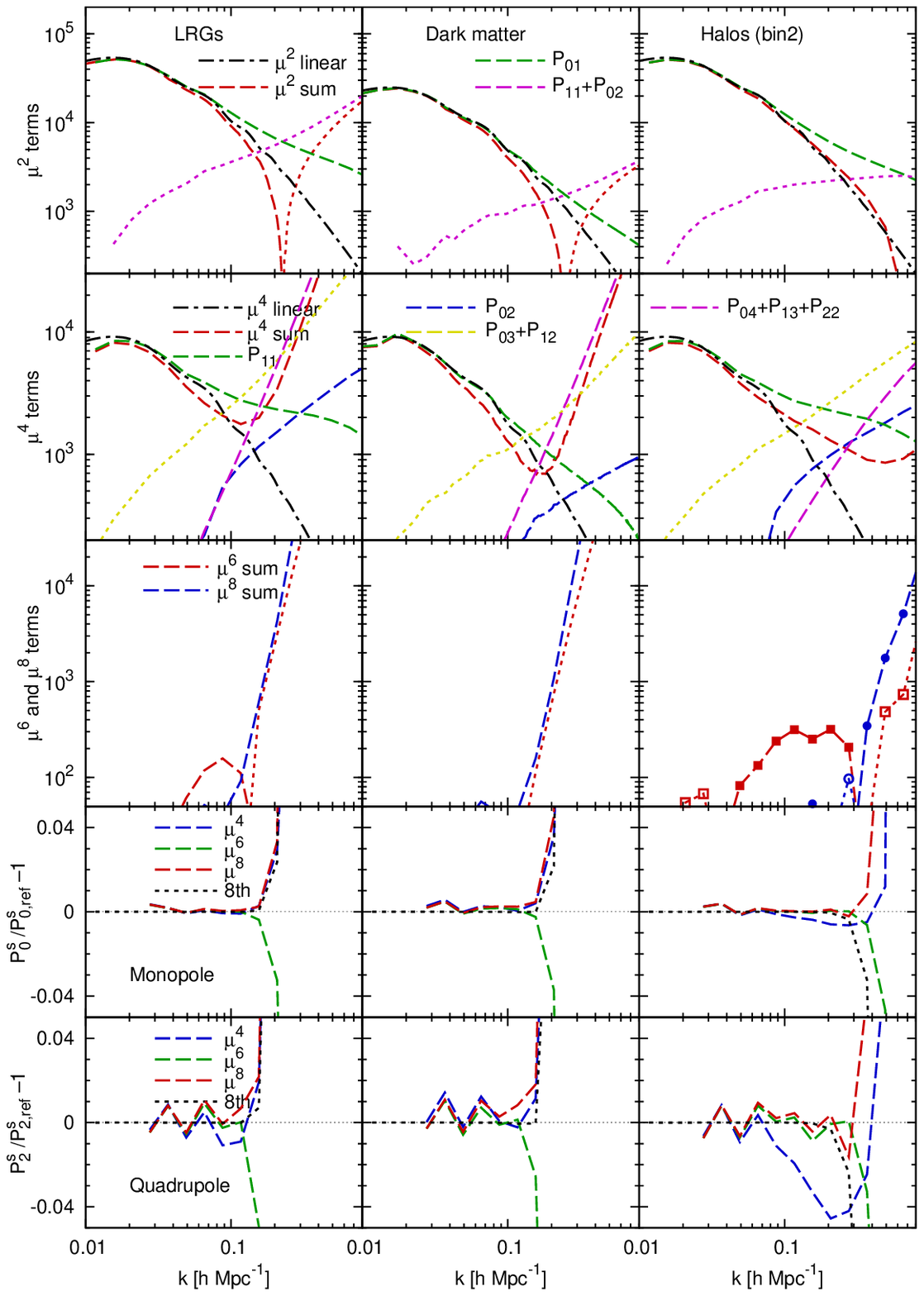}}
\caption{Same as figure \ref{fig:mu_z007} but at $z=0.5$ for LRGs
  (left) and dark matter (middle).  The result for halos with the same
  bias as LRGs (bin2), already presented in figure \ref{fig:mu_z005},
  is shown again at the right panels for comparison.  }
\label{fig:mu_lrg}
\end{figure}
%%%%%%%%%%%%%%%%%%%%%%%%%%%%%%%%%%%%%%%%%%%%%%

In this section we determine the coefficients of $\mu^{2j}$ for each
$P_{LL'}$.  Because we know the exact angular dependence of each
$P_{LL'}$'s \cite{Seljak:2011} (see section \ref{sec:th_angular}), we
treat the coefficients as free parameters and compute the $\chi^2$
statistics for the measurement of each $P_{LL'}$.  We present the
results of the expansion in terms of $\mu^{2j}$ and the contributions
to them from each $P_{LL'}$ up to 8th order in figures
\ref{fig:mu_z007} -- \ref{fig:mu_lrg}.  Those for redshifts $z=0$,
$0.5$, 1 are shown in figures \ref{fig:mu_z007}, \ref{fig:mu_z005} and
\ref{fig:mu_z004}, respectively. The results at $z=2$ for halos of bin1
are also shown in figure \ref{fig:mu_z004}.  We show the comparison of
the results at $z=0.5$ for LRGs, dark matter and halos with the same
bias as LRGs (bin2) in figure \ref{fig:mu_lrg}.  We discuss the
results in detail below.

\subsection{$\mu^0$ term}
The lowest order, $\mu^0$ term in the power spectrum is the real space
power spectrum $P_{00}$.  It is shown for dark matter, LRGs and halos
with one mass bin at $z=0.5$ in figure \ref{fig:pkmu_individual_lin}.
The $\mu^0$ term has the dominant contribution to the monopole, as can
be seen from figure \ref{fig:pkmu_individual_lin}.

\subsection{$\mu^2$ terms}
There are three terms contributing to the coefficient of the $\mu^2$
term in $P^s$, $P^s_{01}$, $P^s_{11}$ and $P^s_{02}$.  At the top
panels in figures \ref{fig:mu_z007} -- \ref{fig:mu_lrg} we show the
individual term contributions to the coefficient of the $\mu^2$ term
in $P^{s}$ as well as the sum.  The linear theory prediction of the
term, $2fb_{1}^{mh}P^m_{00,{\rm lin}}(k)$, is also plotted as black
dot-dashed line, with $b_{1}^{mh}$ is a constant computed from the
mass-halo cross power spectrum at large scales and shown in table
\ref{tab:halo}.  For $\mu^2$ $P^s_{01}$ dominates for low $k$, as that
is the only term which does not vanish in linear theory.  This term
follows linear theory prediction for low $k$, while for $k>0.1\hmpci$
it exceeds the linear theory, more so for the more biased halos.

%%%%%%%%%%%%%%%%%%%%%%%%%%%%%%%%%%%%%%%%%%%%%%
%Figure 10
\begin{figure}%[!t]
\subfigure{\includegraphics[width=1.\textwidth]{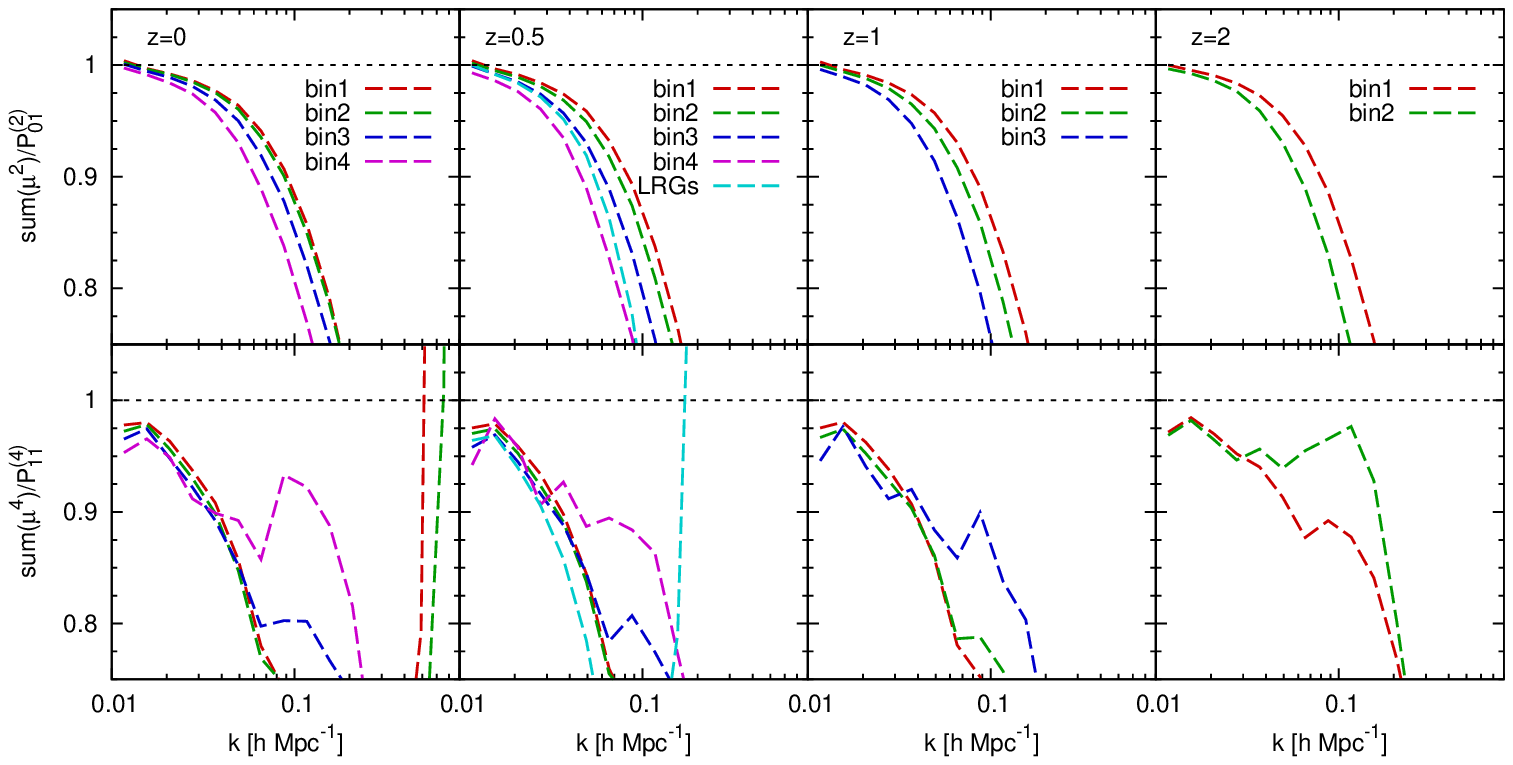}}
\caption{ Contributions to $\mu^2$ (upper panels) and $\mu^4$ (lower
  panels) terms in $P^s$ divided by the dominant terms, $P^{(2)}_{01}$
  and $P^{(4)}_{11}$, respectively.  }
\label{fig:mu_sum}
\end{figure}
%%%%%%%%%%%%%%%%%%%%%%%%%%%%%%%%%%%%%%%%%%%%%%

This is partially canceled by $P^s_{11}+P^s_{02}$.  The $P^s_{02}$
term is negative on small scales and the $P^s_{11}$ term is always
positive, but the sum of the two is negative.  As discussed above
\cite{Seljak:2011}, the sum of the two cancels all the bulk motion
contributions, including any possible shot noise.  At the top panels
of figures \ref{fig:mu_z007} -- \ref{fig:mu_lrg} the sum of the
$P^s_{11}$ and $P^s_{02}$ terms is shown for the coefficient of
$\mu^2$.  Although each of them has a high amplitude (figure
\ref{fig:pkmu_individual_lin}), the sum for halos is suppressed and
lower than that for dark matter because the velocity dispersion of
halos is essentially zero.  To see this cancellation more clearly, we
show in the upper panels of figure \ref{fig:mu_sum} the total
contributions to $\mu^2$ term divided by the dominant term $P_{01}$.
In each case we have about 10\% suppression of $P_{01}^s$ at $k \sim
0.1\hmpci$.

As one can see in figure \ref{fig:mu_lrg}, the total sum of the
$\mu^2$ terms for the dark matter becomes negative due to a large
$P^s_{02}$ caused by small scale velocity dispersion. As expected, the
same behavior can be seen for LRGs in the second panels from the left
in figure \ref{fig:mu_lrg}.  We do not see such behaviors for halos as
in figures \ref{fig:mu_z007} -- \ref{fig:mu_z004}.  The result for
halos with the same bias as LRGs at $z=0.5$ (bin2) is plotted again at
the right panels of figure \ref{fig:mu_lrg} for comparison.

\subsection{$\mu^4$ terms}
The coefficient of the $\mu^4$ term in the power spectrum $P^s(k,\mu)$
contains contributions from 7 different terms, $P_{11}^s$, $P_{02}^s$,
$P_{03}^s$, $P_{12}^s$, $P_{04}^s$, $P_{13}^s$ and $P_{22}^s$.  They
are shown in the second panels of figures \ref{fig:mu_z007} --
\ref{fig:mu_lrg}, together with the summation of the terms.
$P^s_{11}$ is the only term which does not vanish in linear theory and
thus dominates on large scales.  This term approaches the linear
theory prediction on large scales and is above that at small scales,
just like in the case of $P_{00}$ in the $\mu^0$ term and $P_{01}$ in
the $\mu^2$ term.  Likewise, the deviation starts at larger scales and
becomes more prominent for more biased halos.  The lower panels of
figure \ref{fig:mu_sum} show the summation of $\mu^4$ terms divided by
the dominant term $P^s_{11}$.  Even on the largest scales probed here
($k\sim 0.01\hmpci$), the higher-order terms do not vanish and
contribute a few percent.

The next order term in significance should be $P^s_{02}$.  This term
thus does not dominate at any scale, as we have also seen in dark
matter clustering in \cite{Okumura:2012} and can be seen in the left
panels of figure \ref{fig:mu_lrg}.  It is negative on all scales and
act as suppression factors on the $\mu^4$ term in $P^s$.  We have
grouped the other terms together such that they cancel the each
other's shot noise.  The last order we need to consider for the
$\mu^4$ term are the fourth order terms, $P_{04}$, $P_{22}$ and
$P_{13}$.  As discussed in \cite{Okumura:2012} the bulk flow part of
$P_{13}$ cancels out that of $P_{04}$ and $P_{22}$. They also have
shot noise effects which are cancelled out when they are summed over.
We thus show the total contribution from the three terms,
$P_{04}+P_{13}+P_{22}$, in figures \ref{fig:mu_z007} --
\ref{fig:mu_lrg}.  It adds power on small scales. Note that the
contribution from the fourth order terms is smaller than that from the
third order $P_{03}$ for halos, while it exceeds it at $k>0.1\hmpci$
for dark matter and LRGs because of the high random velocities.

\subsection{$\mu^6$ and $\mu^8$ terms}
At order higher than $\mu^4$ we do not have any linear order
contributions, so these terms are expected to be small on large
scales. There are many terms that contribute, third to sixth order
terms in terms of $L+L'$ to $\mu^6$ and fourth to eighth order terms
to $\mu^8$. Third row in figures \ref{fig:mu_z007} -- \ref{fig:mu_lrg}
shows the total contribution from these terms to $\mu^6$ and $\mu^8$
terms. We can see that these terms are indeed negligibly small at
large scales. At smaller scales these contributions increase with the
scale dependences of $k^6$ and $k^8$, respectively.  As $\mu^4$, the
behaviors of $\mu^6$ and $\mu^8$ terms for LRGs are large and similar
to the dark matter, a consequence of non-zero small scale velocity
dispersion.

\subsection{Comparison of $\mu^2$ expansions with Legendre expansions}

It is worth studying how much is a given multipole moment affected by
the expansion in powers of $\mu^2$: we expect this expansion to break
down at high $k$, while at low $k$ it should be strongly convergent.
We transform the redshift-space power spectrum with $\mu^j$ expansions
to Legendre moments using equation (\ref{eq:mu_to_legendre}) and
compare to moments measured directly in redshift space $P^s_{l,{\rm
    ref}}$.

The error for the reconstructed monopole, $P^s_{0}/P^s_{0,{\rm
    ref}}-1$, is shown at the fourth row of figures \ref{fig:mu_z007}
-- \ref{fig:mu_lrg}, while that for the quadrupole,
$P^s_{2}/P^s_{2,{\rm ref}}-1$, is shown at the bottom row.  The blue,
green and red curves show the results when $\mu^{j}$ moments,
$P_{\mu^j}$, are summed up to $\mu^4$, $\mu^6$ and $\mu^8$ terms,
respectively in equation (\ref{eq:mu_to_legendre}).  One can see that
the summation up to $\mu^4$ terms already has a good accuracy at
$k\leq 0.4 \hmpci$ and $k\leq 0.2\hmpci$ for the least massive and
most massive halos at $z=0$, respectively.  However, a more careful
look at the figure reveals that the monopole estimated from the
$\mu^j$ expansion up to $\mu^4$ terms starts to deviate from the
reference spectrum at relatively large scales. The small deviation is
eliminated by adding the $\mu^6$ and $\mu^8$ terms, which means these
high-order nonlinear terms play an important role if one wants to
predict the redshift-space power spectrum accurately.  For the
quadrupole moment, the true power spectrum cannot be reconstructed at
1\% accuracy by using only the terms up to $\mu^4$ even at
$k<0.1\hmpci$, at least for $z=0$. Contributions from $\mu^6$ terms
are necessary for the precise modeling of the quadrupole spectra.
Adding $\mu^8$ terms further improves the accuracy. The deviations of
the reconstructed quadrupole from the true one at large scales are due
to the sampling variance.

For simulated LRGs at $z=0.5$ the convergence is worse: we find that
the series breaks down already at $k \sim 0.1\hmpci$, compared to $k
\sim 0.2-0.3\hmpci$ for the halos of the same bias and redshift. This
is because small scale velocity dispersion inside halos increases the
typical velocity. We also find that including terms above $\mu^4$ does
not improve the convergence: a different approach, such as FoG
resummation is needed in this case \cite{Okumura:2012}.

We also calculate the summed power spectrum, equation
(\ref{eq:p_ss_halo}), which was analyzed for dark matter in detail by
\cite{Okumura:2012}.  We show the accuracy of the summed power
spectrum up to 8th order relative to the reference power spectrum,
shown in the fifth and bottom rows of figures \ref{fig:mu_z007} --
\ref{fig:mu_lrg} for the monopole and quadrupole spectra,
respectively.  We see that the results are similar to the $\mu^8$
results, with no obvious advantage of one over the other.

%%%%%%%%%%%%%%%%%%%%%%%%%%%%%%%%%%%%%%%%%%%%%%
% Section 6 Configuration-space analysis
%%%%%%%%%%%%%%%%%%%%%%%%%%%%%%%%%%%%%%%%%%%%%%

\section{Configuration-space analysis}\label{sec:tpcf}
\begin{figure}%[!t]
\subfigure{\includegraphics[width=1.\textwidth]{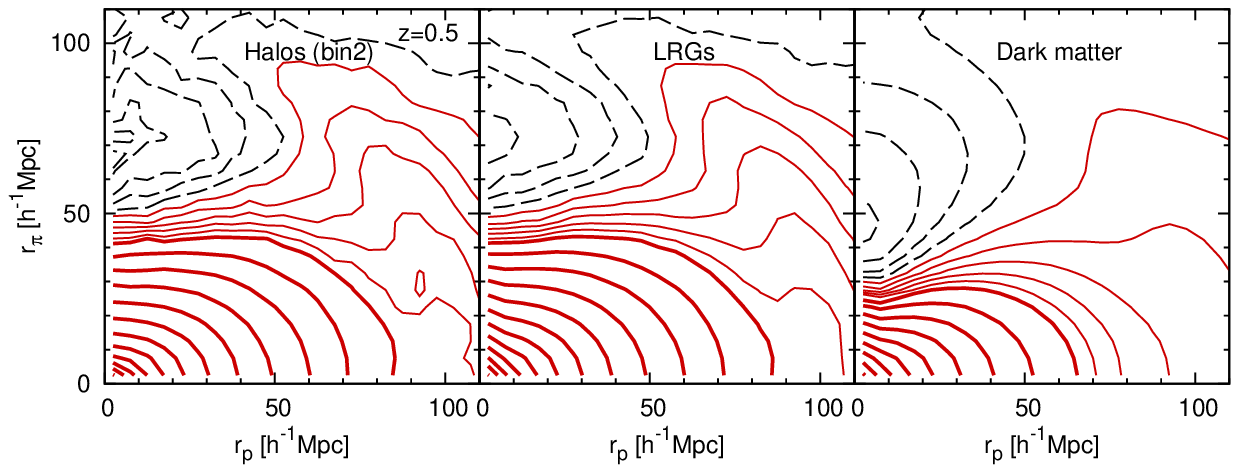}}
\subfigure{\includegraphics[width=1.\textwidth]{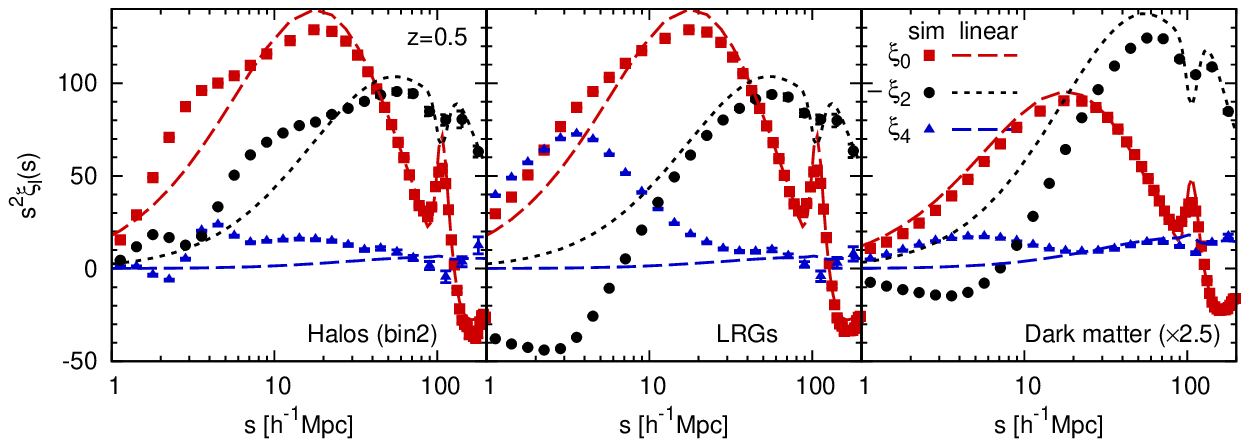}}
\caption{Top: 2D redshift-space correlation functions $\xi(r_p,r_\pi)$
  at $z=0.5$.  Bottom: multipoles of the correlation function.  From
  left to right panels, we show the results for halos (bin2), LRGs,
  and dark matter.  The amplitude of the multipoles of dark matter is
  multiplied by 2.5 for clarity.  }
\label{fig:xi1d_z005_lrg}
\end{figure}

So far our analysis was performed in Fourier space.  In this section
we present the redshift-space correlation function of halos and LRGs
and compare to the power spectrum analysis presented in sections
\ref{sec:power_spectrum} and \ref{sec:legendre}.  Redshift-space
correlation functions are computed as functions of separations
perpendicular ($r_p$) and parallel ($r_\pi$) to the line of sight,
$\xi^{s}(r_p,r_\pi)$.  We show the 2D redshift-space correlation
functions of dark matter, LRGs, and halos with the same bias as the
LRGs, at $z=0.5$ at the top set of figure \ref{fig:xi1d_z005_lrg}.
Since anisotropy caused by the linear Kaiser effect is characterized
by $\beta=f/b$ and $b=1$ for dark matter, the squashing along the line
of sight is more prominent in dark matter clustering than in that of
biased tracers.  On the other hand, on smaller scales nonlinear random
velocities smear the clustering along the line of sight, known as the
FoG effect. Since dark matter and satellite LRGs have larger
velocities, the correlation function at small scales is more elongated
along the line of sight tan the halo correlation.

%%%%%%%%%%%%%%%%%%%%%%%%%%%%%%%%%%%%%%%%%%%%%%
%Figure 11
%%%%%%%%%%%%%%%%%%%%%%%%%%%%%%%%%%%%%%%%%%%%%%

%%%%%%%%%%%%%%%%%%%%%%%%%%%%%%%%%%%%%%%%%%%%%%
%Figure 12
\begin{figure}[!t]
\subfigure{\includegraphics[width=1.\textwidth]{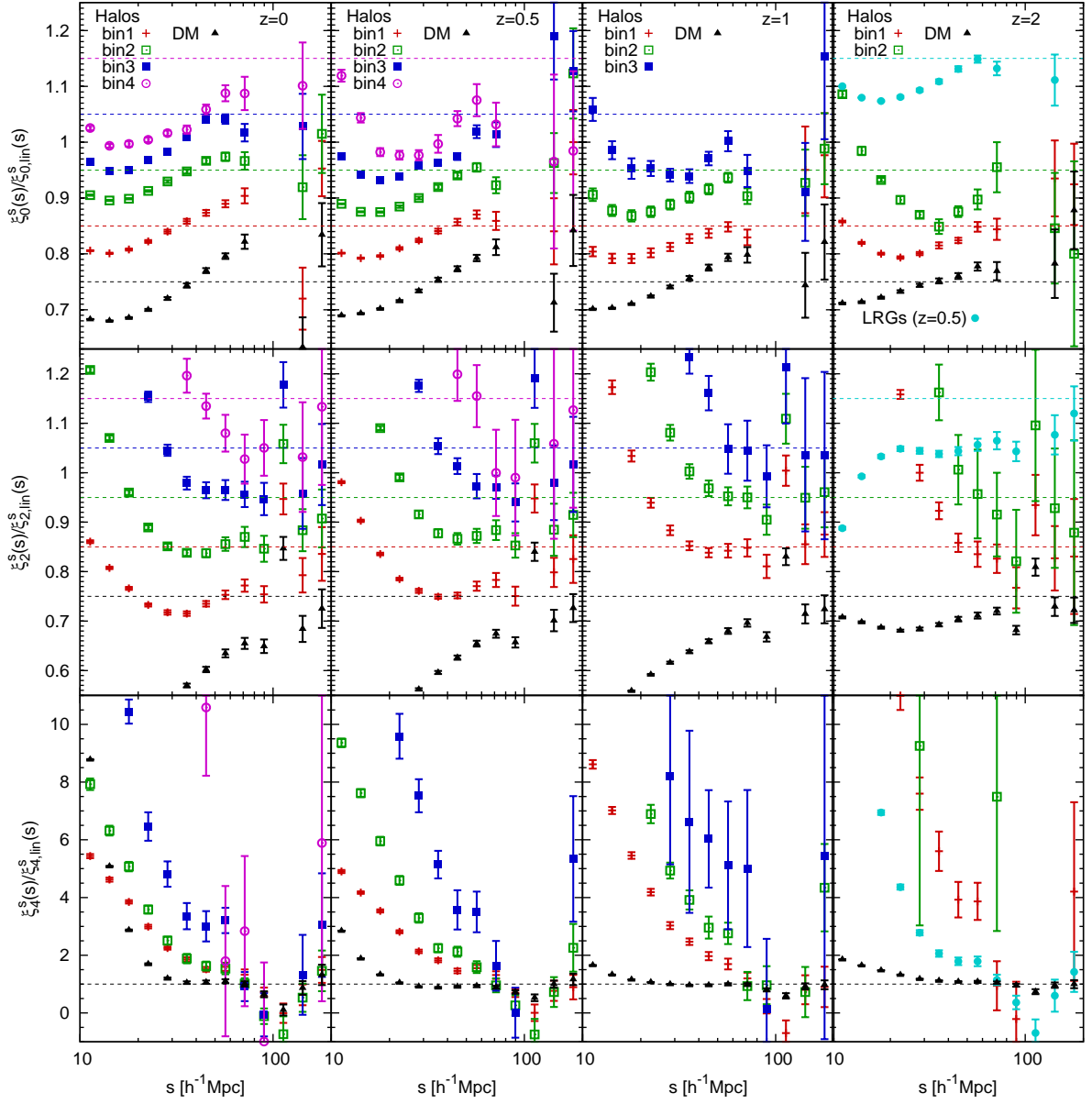}}
\caption{Ratios of multipoles to the corresponding linear theory
  predictions for dark matter, halos and LRGs.  The results for all
  LRGs at $z=0.5$ are shown in the right panels, together with $z=2$
  halos.  For clarity the results of the monopole and quadrupole is
  offset by $-25\%$ for dark matter, by $-15$, $-5$, $+5$ and $+15\%$
  for halos from the lightest to heaviest, and by $+15\%$ for LRGs.
  Two data points of the monopoles at the separation around
  $s=100\himpc$ are not shown because the monopole correlation
  function crosses zero at such scales.  }
\label{fig:xi2d_multi}
\end{figure}
%%%%%%%%%%%%%%%%%%%%%%%%%%%%%%%%%%%%%%%%%%%%%%

The two-dimensional correlation functions can be expanded in terms of
Legendre multipole moments $\xi^s_l$, similarly to the power spectra,
as
\begin{equation}
  \xi^{s}(r_p,r_\pi)=\sum_{l=0,2,4,\cdots}\xi^{s}_l(s){\cal P}_l(\mu) ~, \ \ \ \ \ \ \ \ 
    \xi^{s}_l(s)=(2l+1)\int^{1}_{0}\xi^{s}(r_p,r_\pi){\cal P}_l(\mu)d\mu ~,
\end{equation}
where $\mu=r_\pi/s$.  The multipole moments of the redshift-space
correlation function, $\xi^s_l$, are related to the Fourier
counterparts, $P^s_l$ (equation (\ref{eq:multipole_k2})) through
\be
\xi^s_l(s) = i^l \int \frac{dkk^2}{2\pi^2} P^s_l(k)j_l(ks). 
\ee
The three lowest-order multipoles which contain linear order
contributions, the monopole, quadrupole, and hexadecapole, for the
same three samples and the corresponding predictions from linear
theory are shown at the bottom panels of figure
\ref{fig:xi1d_z005_lrg}.  For the linear theory predictions for LRG
and halos, the constant bias $b_1$ as measured in Fourier space using
the cross-power spectrum (Table \ref{tab:halo}) has been used.  The
amplitude of the multipole moments for dark matter is multiplied by
2.5 for clarity.  The monopole for the dark matter is below the linear
theory, similar to the Fourier space result, while for LRGs and halos
the monopole is above linear theory below $10\himpc$, also similar to
the Fourier space analysis where the transition happens at $k \sim
0.1\hmpci$.

The quadrupole for the dark matter deviates from the linear theory on
a much larger scale, and becomes negative for $s< 8\himpc$.  Same
happens for the LRG sample, which contains small scale nonlinear
random velocities from satellites. The halo sample does not show this
and the quadrupole is above the linear theory below 20$\himpc$.  The
hexadecapole shows very large deviations from linear theory, as
expected also from the Fourier analysis.

In figure \ref{fig:xi2d_multi} we show the multipoles of dark matter,
halos and LRGs at all the four redshifts divided by the corresponding
linear theory predictions for $10<s<200\himpc$.  Because the monopole
changes sign at $s\approx 130\himpc$ and the ratios with linear theory
become noisy, we exclude two data points around the scale.  The ratio
of the monopole to the linear theory typically increases with scale
above 10$\himpc$, and is below the linear theory for $s<30-50 \himpc$
\cite{Okumura:2011,Reid:2011a}.  The quadrupole for halos is below the
corresponding linear theory prediction even on very large scales, by
10\% at $s\simeq 90\himpc$.  The quadrupole for dark matter and LRGs
is suppressed compared to the linear theory on small scales due to
virialized random velocities.  The hexadecapole starts to deviate from
the linear theory predictions at very large scales, more so for more
biased halos \cite{Reid:2011a}. The dark matter hexadecapole is also
above the linear theory, but the amplitude of enhancement is much
smaller.  Overall the configuration space analysis is similar and
consistent with the power spectrum analysis, however the nonlinear
and/or scale dependent bias effects are even stronger and in some
cases we do not converge to linear theory even on very large scales.

%%%%%%%%%%%%%%%%%%%%%%%%%%%%%%%%%%%%%%%%%%%%%%
% Section 7 Conclusions
%%%%%%%%%%%%%%%%%%%%%%%%%%%%%%%%%%%%%%%%%%%%%%

\section{Conclusions}\label{sec:conclusion}

The promise of galaxy redshift surveys is that it contains much of the
cosmological information needed to extract information on cosmological
parameters such as the dark energy properties and neutrino
mass. Redshift space distortions can provide important information, in
the sense of tracing the dark matter velocity field, but can also
damage it, in the sense of being responsible for nonlinear effects
that spoil the comparison between observations and linear theory
predictions.

In the phase-space distribution function approach redshift-space
distortions (RSD) can be written as a sum over density-weighted
velocity moment correlators \cite{Seljak:2011}.  In this paper we
extend the previous work and test this approach to RSD of discrete
objects such as dark matter halos and galaxies, which, unlike the dark
matter, are the observables in redshift space.  For this purpose we
construct a large set of cosmological $N$-body simulations, dividing
each dark matter halo catalog into the mass bins for redshifts $z=0$,
0.5, 1 and 2.  As an example of a more realistic galaxy sample we
construct a mock BOSS-type LRG sample by applying the HOD modeling 
to the simulated halos at $z=0.5$.

Because RSD can be expressed as a sum over number weighted velocity
moment correlators, we can individually compute each correlator term
in the expansion of the redshift-space power spectrum and compare to
its dark matter analog. In doing so we thus construct a number of
generalized bias parameters, of which the first three can also be
compared to the linear theory predictions.  We find that velocity
moment correlators deviate from linear theory predictions on a scale
larger than the density-density correlation, more so if the underlying
tracer is strongly biased.  This can be understood by the fact that
the number or mass weighting of the velocity field in these moments
gives rise to scale dependent bias effects even if the density bias
itself is not scale dependent. These effects typically enhance RSD on
small scales for biased tracers relative to that of the dark matter.

In addition to these biasing effects one must also include the higher
correlators, which are often described as a Fingers of God effects.
It is often assumed that RSD are suppressed due to these higher order
effects, which can lead to smearing of galaxies in the radial
direction.  We find that this effect is small for halos, which are not
sensitive to the small scale velocity dispersion, and as a result RSD
in halos are enhanced at high wavenumber relative to linear theory
predictions, contrary to the commonly assumed model. This changes once
a more realistic galaxy sample with satellites is analyzed, since
satellites provide small scale velocity dispersion, but the amplitude
of the effect depends on how these galaxies are populated inside the
halo and what is the satellite fraction. It may be possible to
construct galaxy samples with a small satellite fraction for which the
small scale velocity dispersion is negligible.

We explore a number of different statistics in this paper: the full
2-dimensional power spectrum $P^s(k,\mu)$, as well as its Legendre
moments monopole, quadrupole and hexadecapole. These all receive an
infinite number of velocity moment correlators and we explore the
convergence.  We also explore expansion in powers of $\mu^{2j}$, each
of which contains a finite number of terms for a given $j$.  We have
computed the coefficients of $\mu^{2j}$ terms of dark matter halos and
LRGs up to $\mu^8$.  Finally, we also explore the configuration space
statistics such as the monopole and quadrupole of the correlation
function, finding good agreement with previous analyses
\cite{Reid:2011a}.

In all cases we find that nonlinear and/or scale dependent bias
effects are very large even on large scales, specially for biased
tracers.  For example, nonlinear and biasing effects for LRGs similar
to those in SDSS are of the order of 10\% at $k \sim 0.15\hmpci$ for
the monopole, $k \sim 0.07\hmpci$ for the quadrupole and even smaller
$k$ in the hexadecapole.  This is a much larger scale (smaller $k$)
than the often stated assumption that they are negligible up to $k
\sim 0.1\hmpci$. This makes a large difference in the amount of
information on $f\sigma_8$ we can extract from RSD, since the signal
scales as $k^{3/2}$, where $k$ is the maximum Fourier wavevector we
can still model.  Some of this can be restored with a nonlinear model
of RSD, but most of the models proposed so far do not account for the
scale dependent bias of $P_{01}$ and $P_{11}$ induced by the $b>1$
galaxies \cite{Scoccimarro:2004, Nishimichi:2011}.  We will present a
perturbation theory based study of the nonlinear and biasing effects
elsewhere (Vlah et. al. 2012, in preparation).

\acknowledgments We would like to thank Pat McDonald, Zvonimir Vlah,
Tobias Baldauf, and Beth Reid for useful discussions.  This research
was supported by the DOE, and the Swiss National Foundation under
contract 200021-116696/1 and Republic of Korea WCU grant R32-10130.
V.D. acknowledges support by the Swiss National Science Foundation.

%\bibliography{rsd}
\bibliography{ms.bbl}
\bibliographystyle{revtex}

\end{document}